\documentclass{emulateapj}
\usepackage{color}\def\red#1{\textcolor{red}{#1}}\def\blue#1{\textcolor{blue}{#1}}\def\green#1{\textcolor{green}{#1}}
\def\red#1{{#1}}\def\blue#1{{#1}}\def\green#1{{#1}}

\newcommand{\feoh}{[{\rm Fe} / {\rm H}]}
\newcommand{\msun}{\, M_\odot}
\newcommand{\mmd}{M_{\rm md}}
\newcommand{\dm}{\Delta_{\rm M}}
\newcommand{\cempnos}{CEMP-no$s$}
\newcommand{\cemps}{CEMP-$s$}

\def\abra#1#2{[{\rm #1}/ {\rm #2}]}
\shorttitle{Early-Age evolution of Milky Way}
\shortauthors{Komiya et al.}

\begin{document}

\title{Early-Age Evolution of the Milky Way Related by Extremely Metal-Poor Stars}

\author{Yutaka Komiya\altaffilmark{1}, Takuma Suda\altaffilmark{2}, Masayuki Y. Fujimoto\altaffilmark{2}}
\altaffiltext{1}{Astronomical Institute, Tohoku University, Sendai, Miyagi 980-8578, Japan}
\altaffiltext{2}{Department of Cosmoscience, Hokkaido University, Sapporo, Hokkaido 060-0810, Japan}

\begin{abstract} 

We exploit the recent observations of extremely metal-poor (EMP) stars in the Galactic halo and investigate the constraints on the initial mass function (IMF) of the stellar population that left these low-mass survivors of $\feoh \lesssim -2.5$ and the chemical evolution that \red{they} took part in.  
   A high-mass nature of IMF with the typical mass $ \simeq 10 \msun$ \red{for the stars of EMP population} and the overwhelming contribution of low-mass members of binaries to the EMP survivors are derived from the statistics of carbon-enriched EMP stars with and without the enhancement of s-process elements (Komiya et al.~  2007, \red{ApJ, 658, 367)}). 
   \blue{We \red{first examine the analysis to confirm their results for various assumptions on the mass-ratio distribution function of binary members. } 
   \red{As compared with the uniform distribution they used, the increase or decrease function of the mass ratio gives a higher- or lower-mass IMF, and a lower-mass IMF results for the independent distribution with the both members in the same IMF, but the derived ranges of typical mass differ less than by a factor of two and overlap for the extreme cases. } }
   \red{Furthermore, we prove} that the same constraints are placed on the IMF from the surface density of EMP stars estimated from the surveys and the chemical evolution consistent with the metal yields of theoretical supernova models.  
   \red{We then} apply the derived high-mass IMF with the binary contribution \red{to show that the observed} metallicity distribution function (MDF) of EMP stars \red{can be reproduced} not only for the shape but also for the number of EMP stars.  
   In particular, the scarcity of stars below $\feoh \simeq -4$ is naturally explained in terms of the hierarchical structure formation, and there is no indication of significant changes in the IMF for the EMP Population.   
   The present study indicates that 3 HMP/UMP stars of $\feoh < -4$ are the primordial stars that were born as the low-mass members of binaries before the host clouds were polluted by their own supernovae.  

\end{abstract}

\section{Introduction}\label{introduction}

To reveal the nature of the extremely metal-poor (EMP) stars in the Galactic halo is the key to the understanding of the formation process of the Galaxy as well as of the mechanism of star formation in the primordial and very metal-poor gas clouds.  
   Because of the very low abundances of iron and other metals, these stars are thought to be survivors from the early days, and hence, are expected to carry the precious information about the early Universe when they were born while they reside in our nearby space.   
   For a past decade, a lot of EMP stars have been discovered by HK survey \citep{Beers92} and Hamburg/ESO (HES) survey \citep{Christlieb01}, which enables us to use halo EMP stars as a probe into the early Universe. 
   The number of known EMP stars exceeds several hundreds even if we limit the metallicity range below $\feoh \lesssim -2.5$ \red{and disclose the metallicity distribution function of these stars} \citep{Beers05a}. 

One of their observed characteristics is very low frequency of stars below the metallicity $\feoh \simeq -4$.  
   Despite that more than $\sim 160$ stars have been registered in the metallicity range of $-4 \lesssim \feoh \lesssim -3$ by high-dispersion spectroscopy \citep[e.g., see SAGA Database;][]{Suda08}, only three stars were found well below this metallicity; two hyper metal-poor (HMP) stars of $\feoh < -5$, HE 0107-5240 \citep[$\feoh = -5.3$;][]{Christlieb02} and HE 1327-2326 \citep[$\feoh = -5.4$;][]{Frebel05}, and one ultra metal-poor (UMP) star of $-5 < \feoh < -4$, HE 0557-4840 \citep[$\feoh = -4.8$;][]{Norris07}.   
   \red{This} has attracted wide interest, in particular, before the discovery of HE 0557-4840 in-between metallicity of $ -5 < \feoh < -4$.  
   \citet{Karlsson05} points out that such a metallicity cut-off can be interpreted as a result of metal spreading process in the stochastic and inhomogeneous chemical-enrichment model. 
   \citet{Karlsson06} then introduce a period of low or delayed star formation due to the negative feedback by the Population III stars, during which metals spread to explain very low iron-abundance of HMP with the carbon yield from rotating stellar models by \citet{Meynet02}.  
   \citet{Prantzos03} argues an early infall phase of primordial gas to alleviate the paucity of low-metallicity stars. 
   \citet{Tumlinson06} adopts a semi-analytic approach for the hierarchical structure formation and presents the model of inhomogeneous Galactic chemical evolution in an attempt of reproducing the statistical features of EMP stars and the re-ionization of the Universe. 
   He addresses the constraints on the IMF of population III stars, arguing high-mass IMF of the mean mass at $\langle M \rangle \simeq 8-42 \msun$. 
   \citet{Salvadori07} also take a similar approach to investigate the chemical evolution of our Galaxy with the mass outflow from mini-halos.  
   In these former works, \red{the low-mass star formation under the metal-deficient condition} is introduced in rather arbitrary ways, and the proper explanation is yet to be devised about the nature and origin of HMP/UMP stars.   

One of the decisive ingredients in studying the structure formation and chemical evolution of Galactic halo is the initial mass function (IMF) of stars in the early days.  
   Most of existent studies have assumed the IMF of EMP stars more or less similar to that of the metal-rich populations except for HMP and UMP stars.  
   From the observations, however, we know that the EMP stars have the distinctive feature that \red{a fair proportion of them show the surface carbon enhancement relative to iron, the proportion by far larger} than the stars of younger populations \citep{Rossi99}. 
   In addition, it is revealed that the carbon-enhanced extremely metal-poor (CEMP) stars are divided into two sub-groups, CEMP-$s$ and CEMP-no$s$ according to the presence and absence of the enhancement of $s$-process elements \citep{Ryan05,Aoki07}. 
   \red{This also forms a striking contrast with the fact that their correspondences among the younger populations, CH stars and Ba stars, are all observed to exhibit the enhancement of s-process elements.  
   Since the EMP survivors are low-mass stars, the enrichment of these elements are expected only through the mass transfer and/or the wind accretion from the AGB primaries in the binaries.}
   Assuming this binary scenario \red{and the same mechanism of carbon enhancement as the stars of younger populations}, \citet{Lucatello05} argue an IMF with the typical mass of $M_{\rm md}\sim 0.79\msun$ for EMP stars from the surplus of \cemps\ stars. 
   Previously, \citet{Abia01} have also asserted an IMF peaking in the intermediate-mass range of $4-8\msun$ for population III stars from the consideration of Galactic chemical evolution with the CN enrichment among the EMP stars.  
   Furthermore, an IMF with $M_{\rm md} \sim 1.7 - 2.3\msun$ has been is discussed for the old halo stars from the MACHO observation in relation to the prospect that the observed micro-lensing may be caused by an alleged population of white dwarfs \citep{Adams96,Chabrier96}.  

In order to use the carbon-enhancement to constrain the IMF, we should properly take into account the evolutionary peculiarity of EMP stars.  
  \red{It is known that for} the stars of $\feoh \lesssim -2.5$, there are two mechanisms of carbon enhancement, while only one mechanism for the stars of younger populations, Pop.~I and II, and also, that a different mode of s-process nucleosynthesis works \citep{Fujimoto00, Suda04,Iwamoto04,Nishimura08}.  
   \red{Applying} these theoretical understandings \red{to the binary scenario}, \citet[][referred to as Paper I in the following]{Komiya07} find that the IMF for EMP stars has to be high-mass with the typical mass of $M_{\rm md} \simeq 10 \msun$ to explain the observed statistic features of both \cemps\ and \cempnos\ stars. 
   \red{In particular, as a consequence, it follows } that the majority of EMP stars, including CEMP stars, were born as the low-mass members of binary systems with the primary stars which have shed their envelope by mass loss to be white dwarfs and have exploded as supernovae.  
   \red{\citet{Tumlinson07a} discuss the binary scenario for HMP stars in the similar way.}

The purpose of this paper is twofold, first to demonstrate the robustness of the high-mass IMF derived in Paper~I, and then to discuss the implications to the formation and early evolution of Galaxy.  
   In the following, we make a distinction between the total assembly of EMP stars that were born in the early Galaxy, including massive stars which were already exploded as supernovae, and the low-mass EMP stars that are still alive in the nuclear burning stages by calling the former ``EMP population" and the latter ``EMP survivors". 
   In deriving the constraints on the IMF of stars for the EMP population, one has to make the assumptions on the binary characteristics, among which the most crucial is the distribution function of mass ratio between the primary and secondary stars in binaries.  
   Paper~I adopts a flat distribution for simplicity.  
   It seems plausible from the observations of the stellar systems of younger populations \citep{Duquennoy91,Mayor92}, and yet, it is true that the mass-ratio distribution is yet to be properly established both observationally and theoretically even for the binaries of younger populations. 
   Several different mechanisms have been proposed for the binary formation, such as the fragmentation during the collapse and the capture of formed stars, and are thought to give different mass-ratio distributions \citep[see also e.g.,][and the references therein]{Goodwin07}. 
   The distribution may increase or decrease with the mass-ratio, or the two stars may form in the same IMF as suggested for the capture origin.  
   In this paper, we examine the dependence of the resultant IMF on the assumed mass-ratio distributions of various functional forms, including the independent coupling of the both stars in the same IMF to demonstrate that the high-mass nature of IMF of EMP population is essentially unaltered.  

The recent large-scaled surveys of EMP stars provide the additional information on the early history of Galactic halo.   
   A fairly large number of known metal-poor stars (144 and 234 stars of $\feoh < -3$ by the HK and HES surveys, respectively) makes it feasible to discuss the metallicity distribution function \citep{Beers05b}.  
   Moreover, the significant coverage of celestial sphere \citep[6900 and $8225 \hbox{ deg}^2$ by the HK and HES surveys, respectively;][]{Christlieb03,Beers05a} allows to consider the total number of EMP survivors in the Galactic halo.   
   We demonstrate that the latter also places an independent constraint on the IMF of EMP population in combination with the metal yields produced by the EMP supernovae if the binary contribution is properly taken into account. 

We then apply the IMF, thud derived, to discuss the chemical evolution in which the stars of EMP population take part.  
   \red{It is shown that} the resultant IMFs can reproduce the number and slope of observed metallicity distribution functions (MDF) for EMP stars, and also, to give an explanation to the scarcity and origin of HMP/UMP stars with the effects of hierarchical structure formation process included.  
   In this paper, we \red{deal only with the iron production by SNe since we are interested in the MDF}, and discuss the basic characteristics of hierarchical structure formation by using simple analytic approximations. 
   \red{\citet{Tumlinson07b} studies the low-mass star formation taking into account the contribution of binary stars, but his approach is different from ours in the uses the hypothesized IMF with the effect of cosmic microwave background.  
   In addition, he considered only the \cemps\ stars, but not \cempnos\ stars or MDF. } 

This paper is organized as follows. 
   In \S 2, we discuss the constraints on the IMF of EMP population from the statistics of CEMP stars and from the total number of EMP survivors in our Galaxies. 
   In \S 3, we investigate the metallicity distribution of EMP stars in Galactic halo with the formation process of the Galaxy taken into account. 
   Then our conclusions follow with discussion of the origin of observed MDF and also of HMP stars.
   In Appendix, we re-discuss the relationship between the number of EMP survivors, estimated from the surveys, and the metal production by the EMP supernovae with the binary contribution taken into account, to demonstrate that they entail the same IMFs as drawn independently from the statistics of CEMP stars. 

\section{Constraints on IMF of EMP stars}\label{IMFs}

In this section, we revisit the problem of constraining the IMF for the stars of EMP population from the observations of EMP survivors, studied in Paper~I.   
   The method is based on the analysis of statistics of CEMP stars in the framework of binary scenario, and hence, involves the assumptions of EMP binary systems. 
   We start with reviewing the method and assumptions used in Paper~I in deriving the constraints on the IMF of EMP population stars.  
   We first investigate the dependence of resultant IMF on these assumptions, in particular of the mass-ratio distribution of binary members.  
   We then discuss the iron production by EMP population stars in relation to the total number of EMP survivors, estimated from the HK and HES surveys, to assess the constraints on the IMF through the chemical evolution of Galactic halo.  

\subsection{Method and basic assumptions}

We give the outline of our method in studying the statistics of CEMP stars and chemical evolution of Galactic halo with the discussion of the assumptions involved, and a brief summary of the observational facts that our study rely on. 

\subsubsection{\red{Statistics of CEMP stars }}

\red{Our method is founded on the results of stellar evolution that the stars of $\feoh \lesssim -2.5$ and of mass $< 3.5 \msun$ undergo hydrogen mixing into the helium convection during the helium core or shell-flashes, differently from the stars of younger populations, Pop.~I and II \citep{Fujimoto90,Fujimoto00}. 
   This triggers the helium-flash driven deep mixing (He-FDDM) to carry out carbon to the surface\citep{Hollowell90}. 
   It is necessarily accompanied with the s-process nucleosynthesis in the helium convection as mixed protons are transformed into neutrons \citep{Suda04,Iwamoto04,Nishimura08}. 
   For EMP stars, He-FDDM works as the mechanism to enrich both carbon and s-process elements in their surface in addition to the third dredge-up (TDU), the latter of which works in the stars of $M > \sim 1.5 \msun$ in common with the stars of younger populations.  }

Consequently, the origins of two sub-groups of CEMP stars are identified with these two mechanisms.  
   The \cemps\ and \cempnos\ stars stem from the low-mass members of EMP binaries with the primaries in the mass ranges of $0.8 \msun < M < 3.5 \msun$ and $3.5 \msun \le M \le M_{up}$, respectively. 
   Here $M_{up}$ is the upper limit to initial mass of stars for the formation of white dwarfs.   
   We take $M_{up} = 6.0 \msun$ \citep[][see also Siess 2007]{Cassisi93}, which is also taken to be the lower mass limit to the stars that explode as supernova.  
   This is the fundamental premise of our study. 
   \red{Among the \cempnos\ stars, there are the stars that show different characteristics such as CS22892-052 with a large enhancement of r-process elements. 
   They may have different origins according to the scenarios such as proposed in connection to supernovae yields \citep[e.g.,][]{Tsujimoto01,Umeda02,Wanajo06}. 
   Accordingly, the \cempnos\ stars may be the admixture of the stars of different origins.  
   Since the ratio between the \cemps\ and \cempnos\ stars depends strongly on the IMF, as shown from Paper~I, however, our results will not affected as long as the \cempnos\ stars contains those with the AGB mass transfer origin that we propose. } 

For the formation of CEMP stars in the binary systems, the initial separation, $A$, has to be large enough to allow the primary stars to evolve through the AGB stage without suffering from the Roche lobe overflow, but small enough for the secondary stars to accrete a sufficient mass of the wind to pollute their surface with the envelope matter processed and ejected by the AGB companion.  
   The lower bound, $A_{\rm min} (m_1, m_2) $, to the initial separation is estimated from the stellar radii of EMP stars taken from the evolutionary calculation \citep{Suda07}, where $m_1$ and $m_2$ are the masses of primary and secondary stars. 
   The AGB star is assumed to eject the carbon enhanced matter of $\abra{C}{H} = 0$ with the wind velocity $v_{\rm wind}=20 \hbox{km s}^{-1}$ until it becomes a white dwarf, and we define CEMP stars as $\abra{C}{Fe} \le 0.5$.  
   The upper bound, $A_{\rm max} (m_1, m_2)$, is estimated by the amount of accreted matter calculated by applying the Bondi-Hoyle accretion rate, 
\begin{equation}
\frac{d m_{2} (t)}{d t} = -\frac{G^{2} m_{2} (t)^{2}}{A(t)^{2} v_{\mathrm{rel}} (t)^{4}} \frac{v_{\rm rel} (t)}{v_{\rm wind}} \times \frac{d m_{1} (t)}{d t}, \label{eq:m1m2} 
\end{equation} 
   in the spherically symmetric wind from the companion, and $v_{\rm rel}$ is the relative velocity of the secondary star to the wind. 
   Accreted matter is mixed in surface convection of depth $0.35 \msun$ and $0.0035\msun$ in mass for giants and dwarfs, respectively. 
   E.g., for the stellar metallicity $\abra{Fe}{H}=-3.5$, the mass of accreted matter has to be larger than $3.5 \times 10^{-4} \msun$ and $3.5 \times 10^{-6} \msun$, and hence, the upper bounds are $\sim 100 {\rm AU}$ $\sim 1000 {\rm AU}$ for dwarfs, respectively.  
   \blue{ \red{ It is pointed out that the} molecular diffusion \citep{Weiss00} and/or the thermohaline mixing \citep{Stancliffe07} work in the envelope of EMP dwarfs to lower the surface abundance of accreted matter by nearly an order of magnitude. 
   If these effects are included, it demands a larger accreted mass, and hence, a decrease of upper bound of binary separation to the carbon enrichment for the dwarf low-mass members with carbon by $\sim 0.5 dex$.   
   This will not affect our results so much since the upper bound is itself sufficiently large to exceed the separation at the peak of period distribution (see below). 
   In addition, it has little effects on the giants which we mainly deal with in the following because of by far deeper surface convection. }
 
If we specify the initial mass function, $\xi_b (m) $, and the distributions of binary parameters, therefore, we can evaluate the frequency of \cemps\ and \cempnos\ stars, and through the comparison with the observations, we may impose the constraints on the IMF and on the binary parameters. 
   The numbers of CEMP-$s$ and CEMP-no$s$ stars currently observable in flux-limited samples 
are given by 
\begin{eqnarray}
  \psi_{{\rm CEMP\hbox{-}}s} & = & f_b \int^{0.8 \, M_\odot}_{ 0.08 \, M_\odot}d m_2 N_s (L[m_2]) \nonumber\\
\times \int_{0.8 M_\odot}^{3.5 M_\odot} &dm_1&  \xi_b (m_1) \frac{n(q)}{m_1}
 \int^{A_M(m_1,m_2)}_{A_{\rm min}(m_1,m_2)} f(P) \frac{dP}{da}da \\
  \psi_{{\rm CEMP\hbox{-}no}s} & = & f_b \int^{0.8 \, M_\odot}_{ 0.08 \, M_\odot}d m_2 N_s (L[m_2]) \nonumber \\
\times \int_{3.5M_\odot}^{M_{up}} &dm_1&  \xi_b (m_1) \frac{n(q)}{m_1}
 \int^{A_M(m_1,m_2)}_{A_{\rm min}(m_1,m_2)} f(P) \frac{dP}{da}da,   
\end{eqnarray}
  where $f_b$ is the binary fraction: 
  $n(q)$ is the distribution of the mass-ratio, $q\equiv m_2/m_1$, and $f(P)$ is the distribution of the period of binaries: and  
   $N(L)$ is the probability of the stars in the Galactic halo with the luminosity $L$ in the survey volume of HES survey.  
\red{   Note that all of them stem from the low-mass members of binary since it is only a very small fraction of stars that have experienced He-FDDM to develop \cemps\ characteristics in themselves and now stay on AGB. } 
   Similarly the total number of EMP survivors 
is given by 
\begin{eqnarray}
&&\psi_{\rm surv}  =  \int^{0.8 \, M_\odot}_{ 0.08 \, M_\odot}dm N_s (L[m]) [ (1-f_b) \xi_s (m) + f_b \xi_b (m) ] \nonumber \\
&&  + f_b \int^{0.8 \, M_\odot}_{ 0.08 \, M_\odot}dm_2 N_s (L[m_2]) \int_{0.8 \, M_\odot }^{\infty} dm_1  \xi_b (m_1) \frac{n(q)}{m_1}, 
\label{eq:surv}
\end{eqnarray} 
\red{   with the contribution of the stars born as single under the initial mass function $\xi_s$). } 
   The rest of the terms give the number of EMP survivors formed as binary, $\psi_{\rm binary}$. 
   \blue{
   The stellar luminosity and lifetime are taken from the evolution calculation of EMP stars by \citet{Suda07}.  
   The AGB primaries dredge up to increase their surface helium abundances, and hence, may cause the surface enrichment of helium to the companion stars in the binaries at the same time with the carbon enrichment, though both suffering the dilution in the envelope convection. 
   The surface enrichment increases the luminosity during the RGB evolution to shorten their RGB lifetime of polluted EMP stars nearly in inverse proportion of the luminosity, but the survey volume increases with the luminosity. 
   For flux limited sample, the observed number of EMP giants may rather increases with the surface enrichment in proportion to a half power under a constant density distribution.  
   \red{On the other hand, the increase in the luminosity occurs only after the hydrogen burning shell comes to take place in the shell to which the pollutants are carried in by the surface convection, and hence, in the later stages for metal-poorer stars. }
   Accordingly, the effect of helium enhancement will little affect our results of constraints on the IMF since HES survey is thought to reach far enough that the spatial distribution of halo stars decreases. 
}

\subsubsection{Model Parameters}\label{subsec-modelpara}

\red{In this paper, we assume that the binary primary stars and single stars are born under the same IMF, i.e., $\xi (m)= \xi_b (m)= \xi_s (m)$. }  
   For the form of IMF, we may well assume a lognormal function with the medium mass, $\mmd$, and the dispersion, $\dm$, as parameters 
\begin{equation}
\xi (m) \propto \frac{1}{m} \exp \left[ -\frac{(\log m-\log \mmd)^2}{2 \dm ^2} \right].    
\end{equation} 
   In addition, we assume the binary fraction $f_b = 0.5$ in this paper. 
   Our results are little affected by the assumption about $f_b$ since not only the CEMP stars but also most of the EMP survivors come from the secondary companions of binaries unless $\mmd < 0.8 \msun$, as seen later.   
   As for the binary period, we may adopt the distribution derived for the nearby stars by \citet{Duquennoy91}, 
\red{
\begin{equation}
f(P) \propto  \frac{1}{P} \exp \left[ \frac{-\left(\log P-4.8\right)^2}{2 \times 2.3^2} \right], 
\end{equation}
   where $P$ is the period in units of days. }
   The binary fractions and period distributions of halo stars are observed to be not significantly different from those of nearby disk stars \citep{Latham02, Carney03}. 
   Additionally, it is shown in Paper~I that this period distribution is consistent with the observations of CEMP stars for the periods of $P \lesssim 10$ yr confirmed to date (see Fig.~3 in Paper~I).   

The mass ratio distribution is an essential factor in discussing the evolution of binary systems, and yet, it is not well understood. 
   The mass ratio distribution of metal-poor halo stars is investigated observationally \citep[e.g., see][]{Goldberg03, Abt08}, and yet, subject to large uncertainties. 
    Especially for the binary with intermediate-mass or massive primary stars, it is hard to know the mass ratio distribution from the observations.  
   Theoretically, neither the fragmentation of gas cloud nor the accretion process onto proto-binaries are yet well understood even for Population~I stars \citep[e.g.][]{Bate97, Ochi05, Machida08}.
   In order to test the assumption on the mass-ratio distribution, we investigate the constraints on the IMF for different mass-ratio distributions.  
   In Paper~I, the simplest flat distribution is assumed in Paper~I among the possible distributions. 
   \blue{In this paper, we test some other assumptions and discuss the dependence of IMF parameters on the mass ratio distributions, as stated in \S\ref{nqsec}. }

We may define the coupling mass distribution function, $\chi (m_1, m_2)$, as the fraction of the binaries with a primary and secondary star in the mass range of $[m_1, m_1 + d m_1]$ and $[m_2,  m_2 + d m_2]$ $(m_1 \ge m_2)$ and write it in the form; 
\begin{eqnarray}
\chi (m_1, m_2)dm_1 dm_2 &=& \xi (m_1) n (q) dq dm_1 \nonumber \\ 
&=&  \xi (m_1) n (m_2, m_1) /m_1 dm_1 dm_2. 
\label{eq:masscouple1}
\end{eqnarray}   
   \red{Here the initial} mass function, $\xi$, of the primary star is assumed to be the same as IMF of single stars:  
    And $n(q)$ is the mass ratio distribution, for which we assume both extremities of increase and decrease functional forms in addition to the constant one, adopted in Paper~I; 
\begin{equation}
n(q) = \left\{
\begin{array} {lc}
1/(1-0.08 \msun/m_1)	& (\rm{Case~A}) \\
2 q / [1 - (0.08 \msun/m_1)^2] & (\rm{Case~B}) \\
q^{-1} / \ln (m_1/0.08 \msun)	& (\rm{Case~C}).  
\end{array}
\right. 
\end{equation}
   Furthermore, we take up a different type of mass-ratio distribution that the primary and secondary stars independently obey the same IMF such as assumed by \citet{Lucatello05}.  
   In this case, the coupling mass distribution function is given as a product of the same IMF as; 
\begin{equation}
\chi(m_1, m_2)dm_1 dm_2 = 2 \xi (m_1) \xi (m_2) dm_1 dm_2  \qquad  ({\rm Case~D}). 
\label{eq:masscouple2}
\end{equation} 
   We shall refer to this distribution function as ``independent" coupling.   
  From the comparison with eq.~(\ref{eq:masscouple1}), we may write the mass-ratio function in the form $n (m_2 , m_1) = 2m_1 \xi(m_2)$;  
   it is should be noted, however, that the frequency of binaries with a primary star of mass $m_1$ is not normalized and increases with $m_1$ from zero to 2, as given by the integral $\int^{ 1}_{0.08\msun/m_1} n(m_1, m_2) d q = 2 \int^{m_1}_{0.08\msun} \xi (m_2) dm_2$. 

With these specification and with the assumed mass-ratio distribution function, we may compute the fractions of EMP survivors, $\psi_{\rm{surv}}(\mmd, \dm)$ and of both cemps\ stars, $\psi_{{\rm CEMP\hbox{-}}s}(\mmd, \dm)$ and $\psi_{{\rm CEMP\hbox{-}no}s}(\mmd, \dm)$, and search the ranges of the IMF parameters, medium mass $\mmd$ and dispersion $\dm$, that can reproduce the statistics of CEMP stars consistent with observations

\red{\subsubsection{Total iron yield of EMP supernovae}}

We can pose another constraint from the total iron yield, $M_{\rm Fe}$, of EMP population and the total number, $N_{\rm EMP, G}$, of the giant EMP survivors.  
   For $N_{\rm EMP, G}$, estimated from the results of existent surveys, the total stellar mass, $M_{\rm EMP}$, of EMP population for an assumed IMF is given by, 
\begin{equation} 
M_{\rm EMP} (\mmd, \dm) = \overline{ m} \ N_{\rm EMP, G} / f_{\rm G}, 
\label{eq:EMP-mass}
\end{equation}
  where $f_{\rm G}$ is the fraction of giant EMP survivors in all the stellar systems, born as EMP population, and $\overline{ m}$ is the averaged mass of EMP population stars: 
\begin{equation}
f_{\rm G} = \left[ \xi (0.8 \msun) + f_b \int_{0.8 \msun} \xi(m_1) n (0.8\msun / m_1) {d m_1 \over m_1} \right] \Delta M_{\rm G}, 
\label{eq:frac-giants}
\end{equation}
\begin{equation}
\overline{ m} = \int dm_1 [ m_1 \xi (m_1) + {f_b \over m_1} \int ^{m_1} m_2 n (q) dm_2 ]. 
\end{equation}
   The first terms of both equations denote the contributions by the stars born as the single stars and as the primary stars in the binaries and the second terms denote the contributions by the stars born as the secondary stars in the binaries. 
   The mass and mass range of EMP stars now on the giant branch are taken to be $M=0.8 \msun$ and $\Delta M_{\rm G} = 0.01 \msun$, based on the stellar evolution calculation of stars with $\feoh = -3$, as in paper~I. 

The massive stars of EMP population have exploded as supernovae to enrich the interstellar gas with metals. 
   The amount of iron, $M_{\rm Fe, EMP}$, ejected by all the supernovae of EMP population of the total mass, $M_{\rm EMP}$, is given by
\begin{equation}
M_{\rm Fe, EMP} = {M_{\rm{EMP}} \over \overline{m}} f_{\rm{SN}} \langle Y_{\rm Fe} \rangle  = N_{\rm{EMP, G}} {f_{\rm{SN}} \over f_{\rm{G}}} \langle Y_{\rm Fe} \rangle, 
\label{eq:EMP-iron}
\end{equation}
   where $f_{\rm{SN}}$ is the fraction of the stars that have exploded as supernovae and given by, 
\begin{equation}
f_{\rm SN} = \int_{M_{up}} dm_1 \xi (m_1) [ 1 + {f_b \over m_1} \int ^{m_1}_{M_{up}} n (q) dm_2 ]:  
\label{eq:frac-SN}
\end{equation}
   and $\langle Y_{\rm Fe} \rangle$ is the averaged iron yield per supernova, taken to be $\langle Y_{\rm Fe} \rangle = 0.07 \msun$ in the following calculations. 

With these evaluations and the observed number of EMP giants, we can give the total iron yield of stars of EMP population as a function of IMF parameters. 
  The comparison with the total amount of iron estimated from the chemical evolution of Galactic halo may impose constraint on the IMF parameters. 

\red{\subsubsection{Observational constraints}}

The first constraint is the number fraction of \cemps\ stars. 
The HK and HES observations tell that the CEMP stars with $\abra{C}{Fe}>1$ account for $20 \sim 25\%$ of EMP stars \citep[e.g.,][]{Beers99, Rossi99, Christlieb03}.   
   \citet{Cohen05} suggest a slightly lower fraction of $14.4\% \pm 4\%$ with the errors in the abundance analysis taken into account while \blue{\citet{Lucatello06} obtain a larger frequency of $21\%\pm 2\%$ for the HERES (HES r-process enhanced star) survey sample, both for $\abra{Fe}{H}<-2.0$. 
  It is claimed that the frequency of CEMP is higher at lower metallicity of $\abra{Fe}{H}<-2.5$ but we have to subtract contribution from the \cempnos\ stars. } 
  In this paper, we adopt the observational constraint on the fraction of the \cemps\ stars at $10-25\%$;
\begin{equation}
0.1 < {\psi_{{\rm CEMP\hbox{-}}s}(\mmd, \dm) \over \psi_{\rm{surv}}(\mmd, \dm)} < 0.25 .
\end{equation}

The second constraint is the number ratio between \cempnos\ and \cemps\ stars. 
   The observed frequency of \cempnos\ to \cemps\ stars is $\sim 1/3$ or more, \citep[e.g.,][]{Ryan05,Aoki07}.  
   \citet{Aoki07} point out that it increases for lower metallicity, reporting the ratio as large as $9/14$ for $\feoh \le -2.5$.   
   In addition, EMP stars enriched with nitrogen are found in number comparable with, or more than, \cempnos\ stars \citep[``mixed'' stars;][]{Spite05}, whose origin can be interpreted in terms of the same mechanism but with more massive primary companions that experience the hot bottom burning (HBB) in the envelope of the AGB. 
   Some other scenarios for CEMP stars have been proposed \citet{Umeda05, Meynet06} but we assume all CEMP stars are formed in binaries with AGB in this paper. 
   We adopt the observational constraint on the relative frequency of \cempnos\ to \cemps\ stars at $1/3-1$; 
\begin{equation}
1/3 < {\psi_{{\rm CEMP\hbox{-}no}s}(\mmd, \dm) \over \psi_{{\rm CEMP\hbox{-}}s}(\mmd, \dm)} < 1. 
\end{equation}

We note that the above two constraints are not dependent on the total mass nor on the spatial distribution of the stellar halo because they are concerned with the relative number ratios. 

The third constraint is the total iron yield from the EMP population. 
   The HES survey obtained 234 stars of $\feoh < -3$ \citep{Beers05b} as a result of the medium-resolution, follow-up observations of 40\% of the candidates, selected by the objective-prism survey of the nominal area $S=8225 {\rm deg}^2$ \citep[][]{Beers05a}.
   Taking the relative frequency between the giants and dwarfs ($1: 0.93$) and the ratio of the stars of $\feoh <-3$ and $\feoh <-2.5$ ($6\% : 20\%$) from their Table 3, we may estimate the total number of EMP giants in the Galactic halo in the flux limited sample at:
\begin{equation}
 \sigma_{\rm{EMP,G}}\simeq 410 
\hbox{ sr}^{-1}.  
\label{eq:HES-G}
\end{equation}
   We assume that all giant stars in the survey areas are observed because of the fairly large limiting magnitude of HES survey ($B=17.5$), about two magnitude deeper than for the HK survey, and neglect the spatial distribution of EMP giant for simplicity since the sufficient information is not yet available (see \S 7.3 in Paper~I for the detail). 
   Then we have the total number of EMP giants $N_{\rm EMP, G} = 5.2 \times 10^3$ in the Galaxy.  

On the other hand, the amount of iron necessary to promote the chemical evolution of the whole gas in Galaxy of mass, $M_h =10^{11}\msun$, up to $\feoh = -2.5$ is as much as 
\begin{equation}
M_{\rm Fe, halo} = M_h X_{\rm Fe, \odot} 10^{-2.5} = 10^{5.5} \msun,   
\label{eq:EMP-ironprod}
\end{equation}  
   and the supernovae of EMP population should have provided this amount of iron unless there were other population(s) of stars which made iron without producing low-mass stars. 
   \red{Using eq.~(\ref{eq:EMP-iron}), this is transferred into a constraint on the IMF as} 
\begin{equation}
M_{\rm Fe, \red{EMP}} = 0.07\msun \times 5.2 \times 10^3 {f_{\rm{SN}}(\mmd, \dm)  \over f_{\rm G}(\mmd, \dm)} \simeq 10^{5.5} \msun.  
\label{eq:const-ironpro}
\end{equation}
   \red{The estimated number of EMP survivors may be subject to significant uncertainties. 
   If we take into account the EMP stars in the outer halo and in the Galactic bulge that HES survey cannot reach, $N_{\rm EMP, G}$ can be larger, which demands a smaller amount of iron produced per EMP survivor, and hence, a smaller number of supernova, leading to a lower-mass IMF. 
   A lower-mass IMF also results if the binary fraction is smaller and/or if there is other source(s) of iron that does not accompany the low-mass star formation. 
   On the other hand, if a part of the supernovae ejecta is dispelled and lost from the Galaxy, it demands a larger amount of iron, $M_{\rm Fe, EMP}$, and hence, a higher-mass IMF.  
   Despite such uncertainties both of the observations and the theoretical assumptions, the constraints on the IMF derived from eq.~(\ref{eq:const-ironpro}) are rather robust since the ratio, $f_{\rm SN}/f_{\rm G}$, is a rapidly varying function of IMF. } 

\subsection{Dependence on Mass-Ratio Distributions}\label{nqsec}

\begin{figure*}
\epsscale{1.0}
\begin{tabular}{cc} 
\begin{minipage}{0.5\hsize}
\plotone{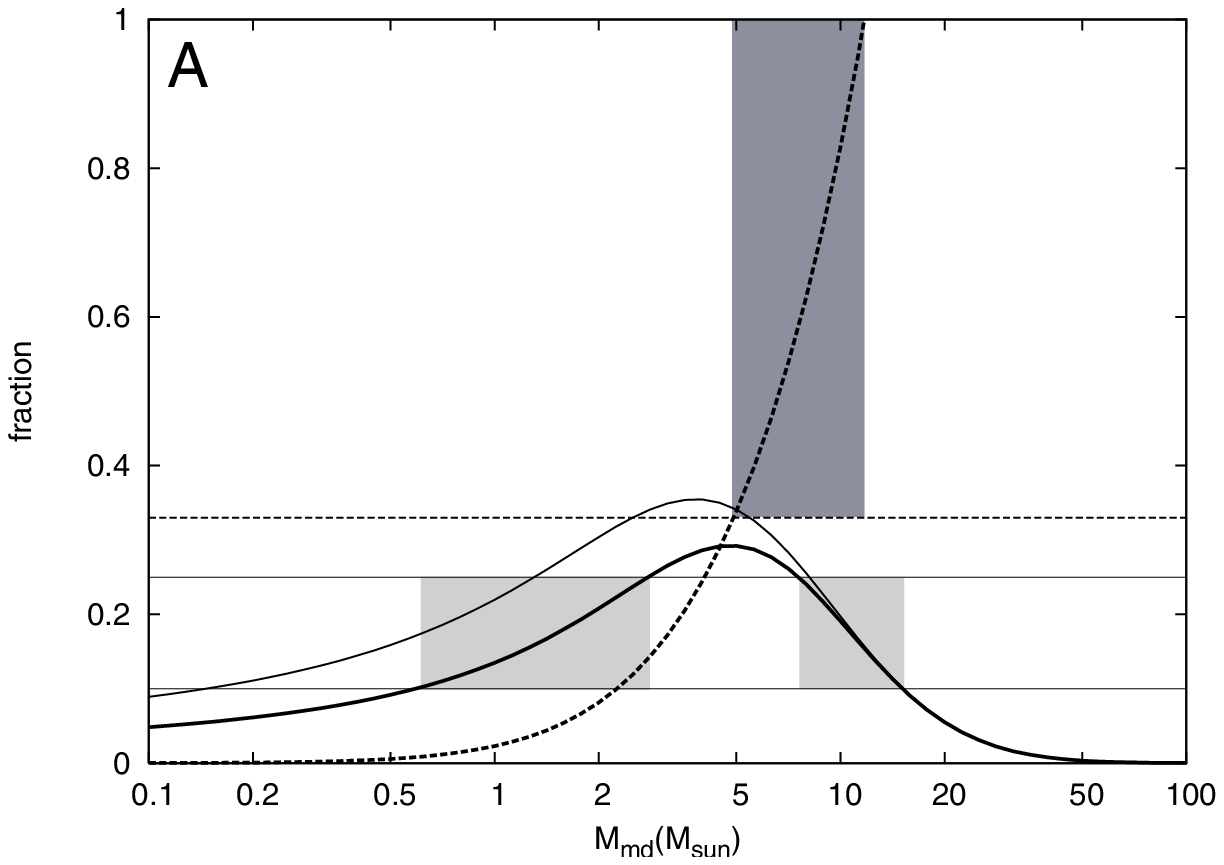}
\end{minipage} 
\begin{minipage}{0.5\hsize}
\plotone{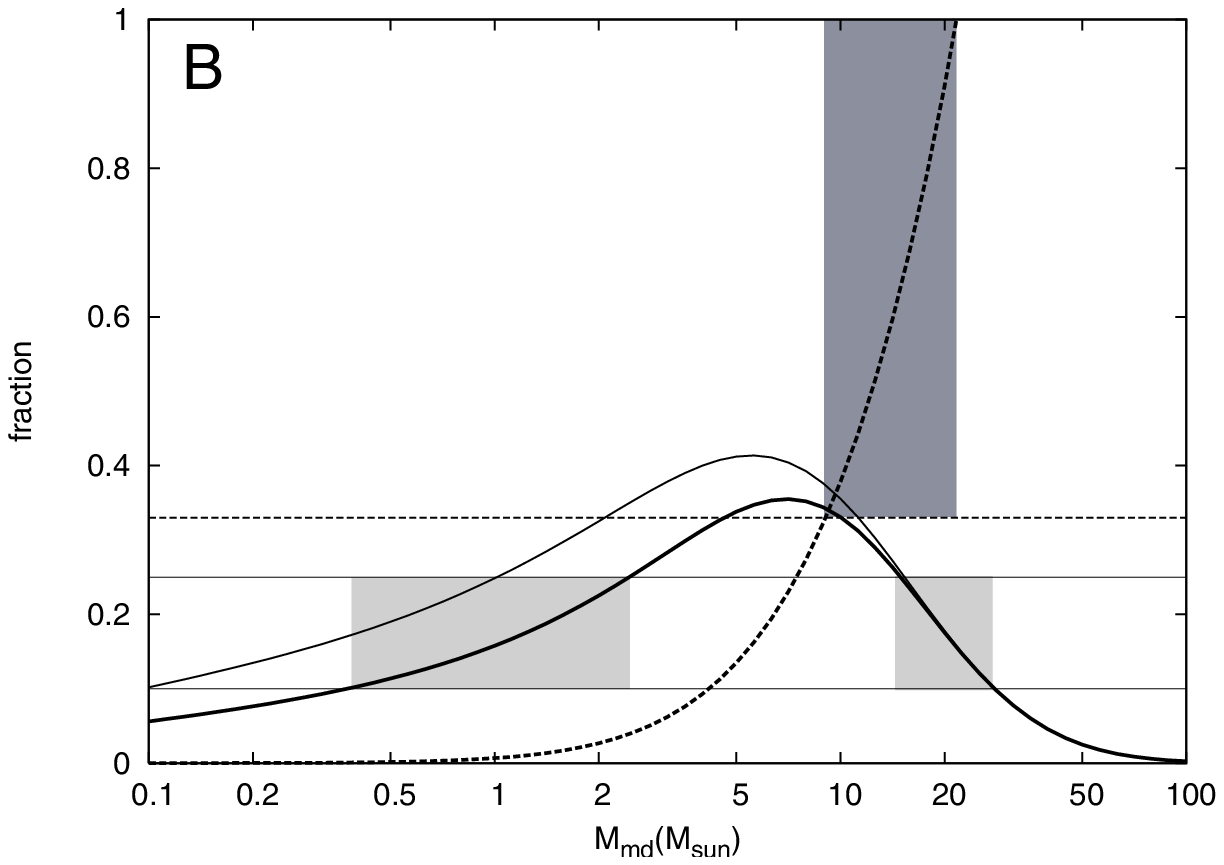}
\end{minipage} \\
\begin{minipage}{0.5\hsize}
\plotone{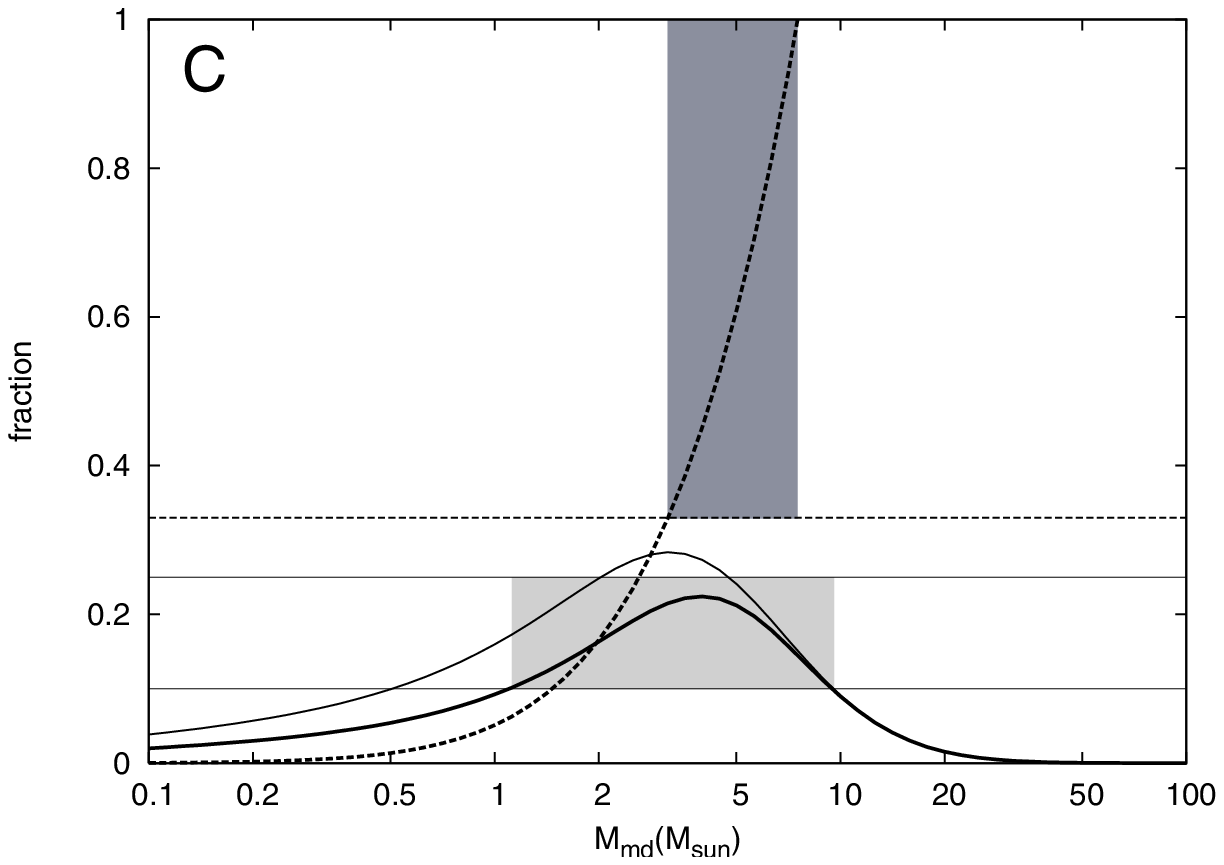}
\end{minipage}
\begin{minipage}{0.5\hsize}
\plotone{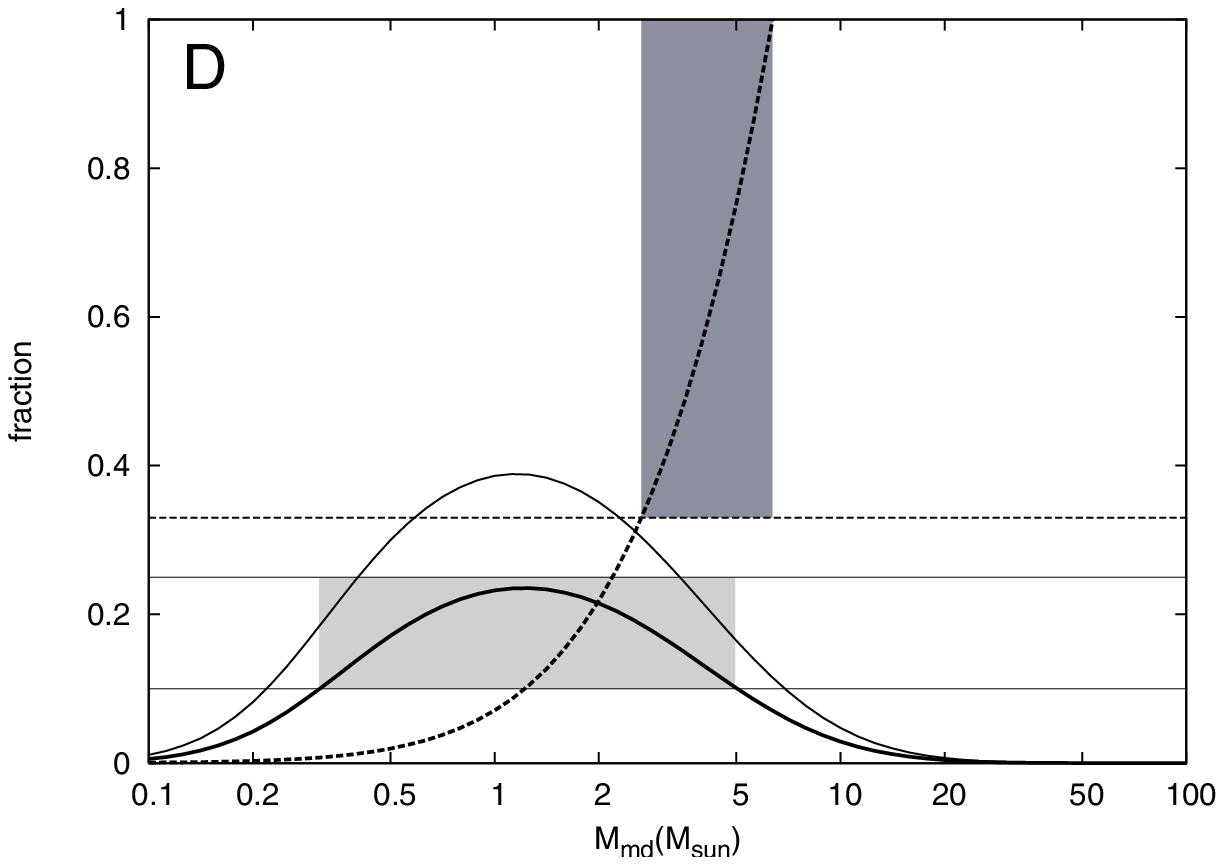}
\end{minipage}
\end{tabular}
\caption{Constraints on the IMF of EMP population for four cases of assumption for mass-ratio distribution function.  
{\it Top left}Case A: $n(q) = const.$, 
{\it Top right}Case B. $n(q) \propto q$, 
{\it Bottom left}Case C. $n(q) \propto 1 / q$, and 
{\it Bottom right}Case D. independent coupling. 
Thin and thick solid lines denote the fraction of CEMP-$s$ in the EMP survivors as the function of the medium mass $\mmd$ for a fixed dispersion of $\dm = 0.33$ for the EMP stars born as binaries ($\psi_{{\rm CEMP\hbox{-}}s}(\mmd, 0.33) / \psi_{\rm{binary}}(\mmd, 0.33)$) and for the total EMP stars including the single stars born in an equal number to the binaries($\psi_{{\rm CEMP\hbox{-}}s}(\mmd, 0.33) / \psi_{\rm{surv}}(\mmd, 0.33)$), respectively. 
  Broken line denotes the ratio of CEMP-no$s$ to CEMP-$s$ stars ($\psi_{{\rm CEMP\hbox{-}no}s}(\mmd, 0.33) / \psi_{{\rm CEMP\hbox{-}}s}(\mmd, 0.33) $). 
  Light and dark shaded areas denote the parameter ranges for the IMFs that can give rise to the observed fraction of CEMP-$s$ stars in the EMP survivors ($10-25\%$) and the observed ratio of CEMP-no$s$ to CEMP-$s$ stars ($1/3-1$), respectively.  
}
\label{fig:psi}
\end{figure*}

\begin{figure*}
\epsscale{1.0} 
\begin{minipage}{0.5\hsize}
\plotone{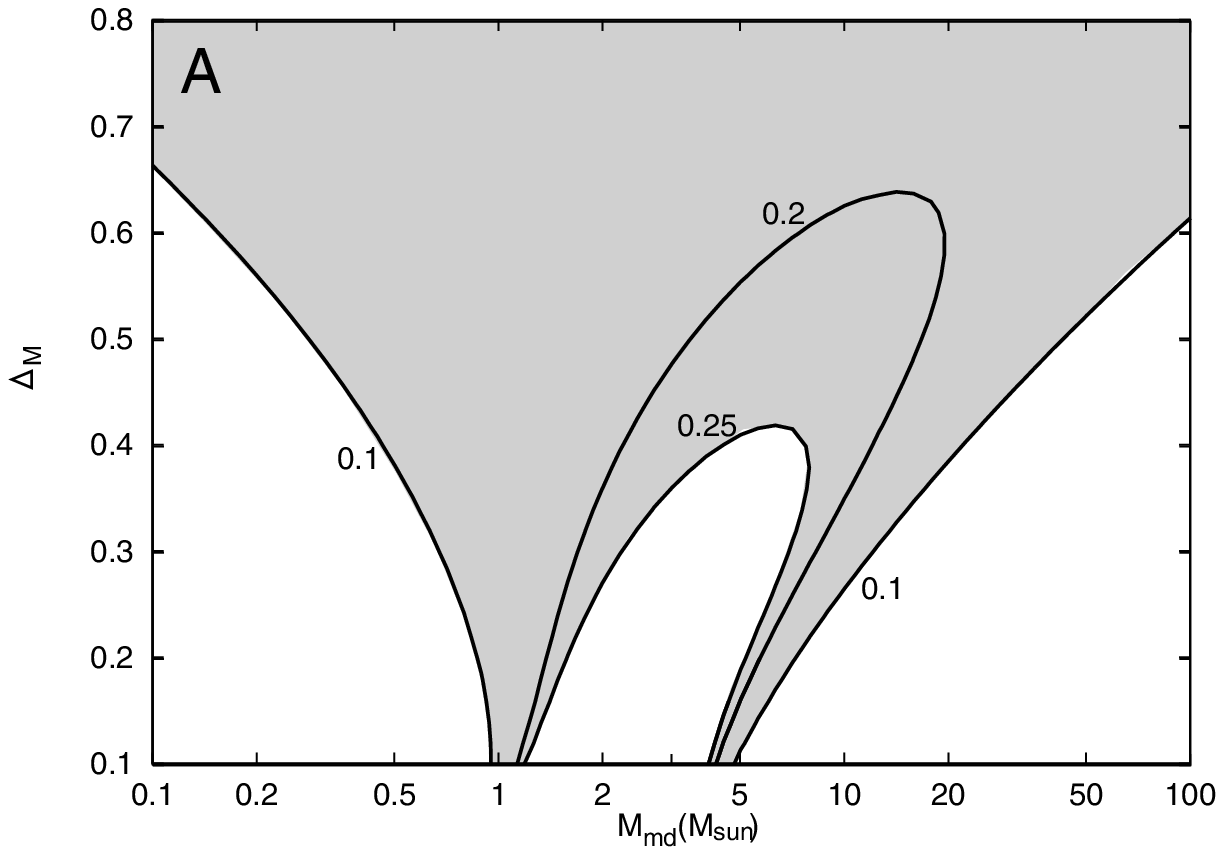}
\end{minipage} 
\begin{minipage}{0.5\hsize}
\plotone{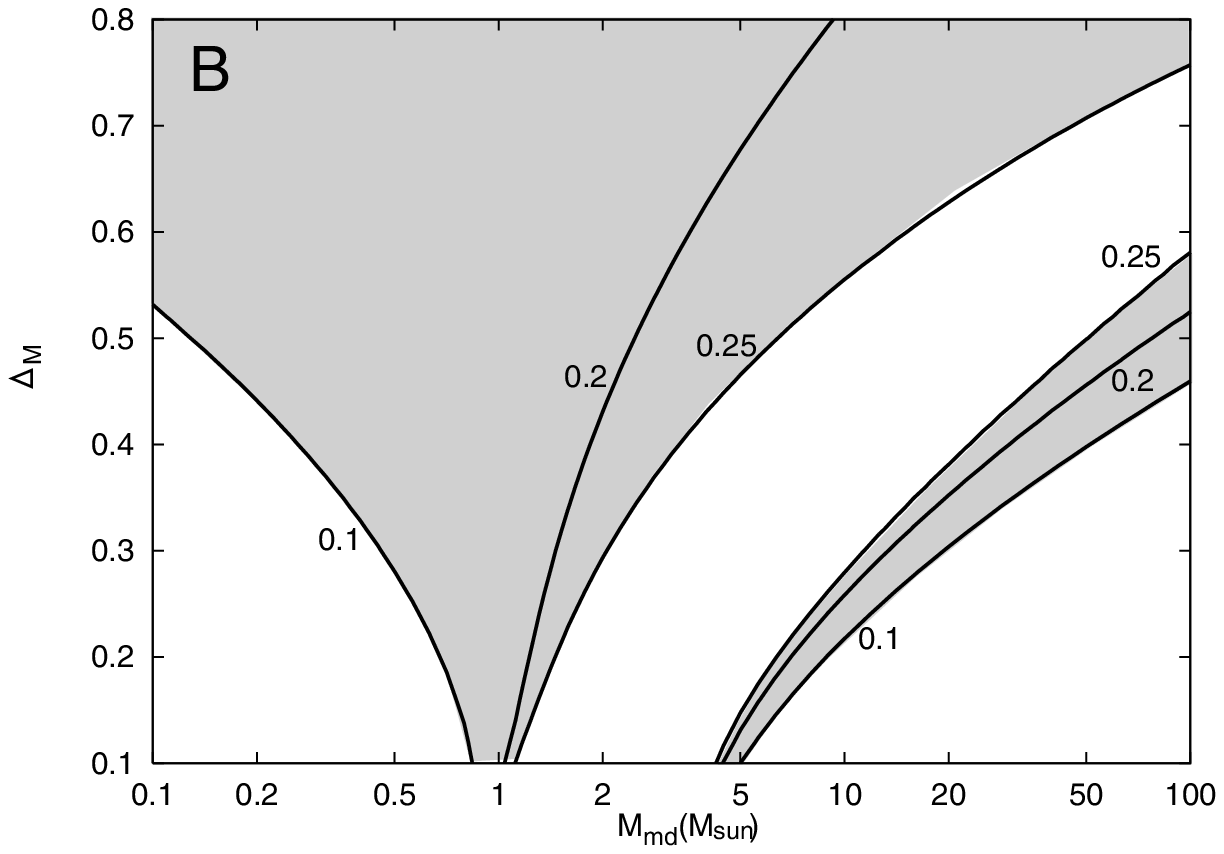}
\end{minipage} 
\begin{minipage}{0.5\hsize}
\plotone{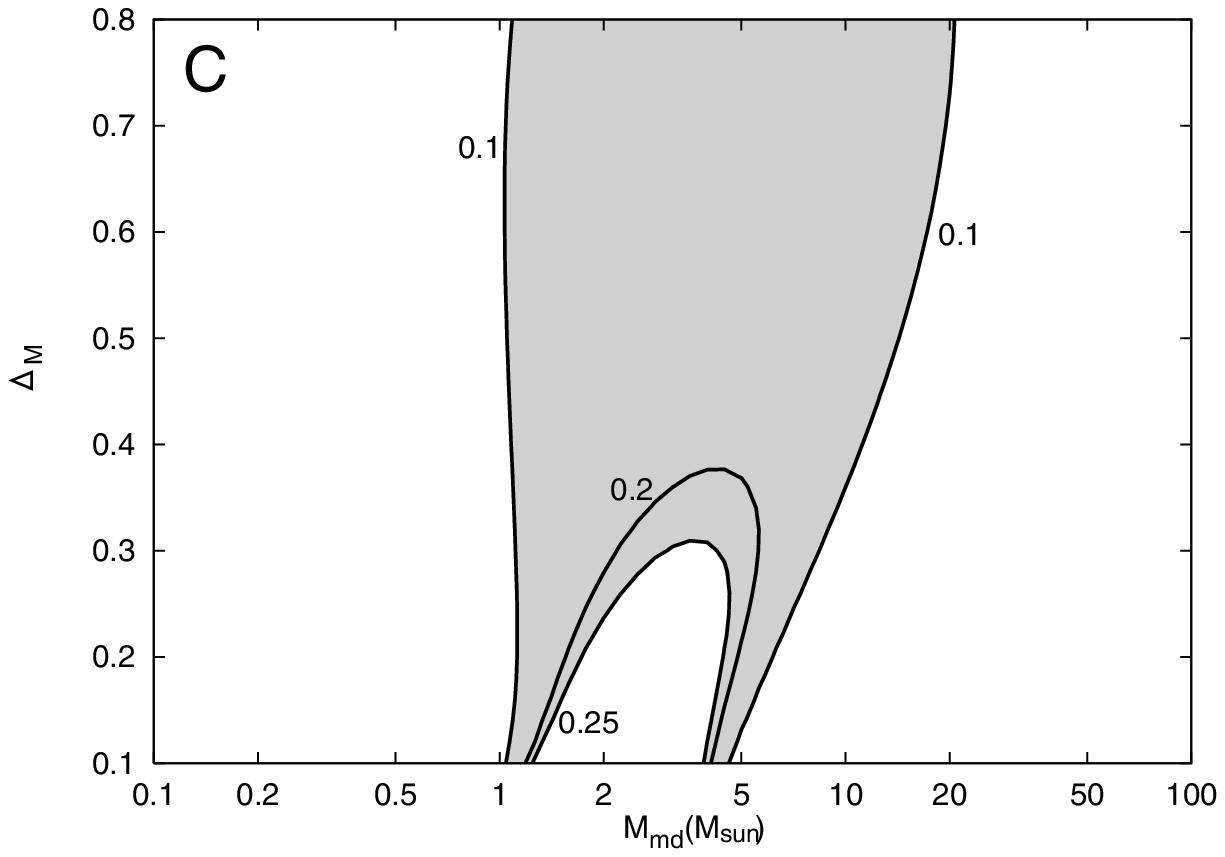}
\end{minipage}
\begin{minipage}{0.5\hsize}
\plotone{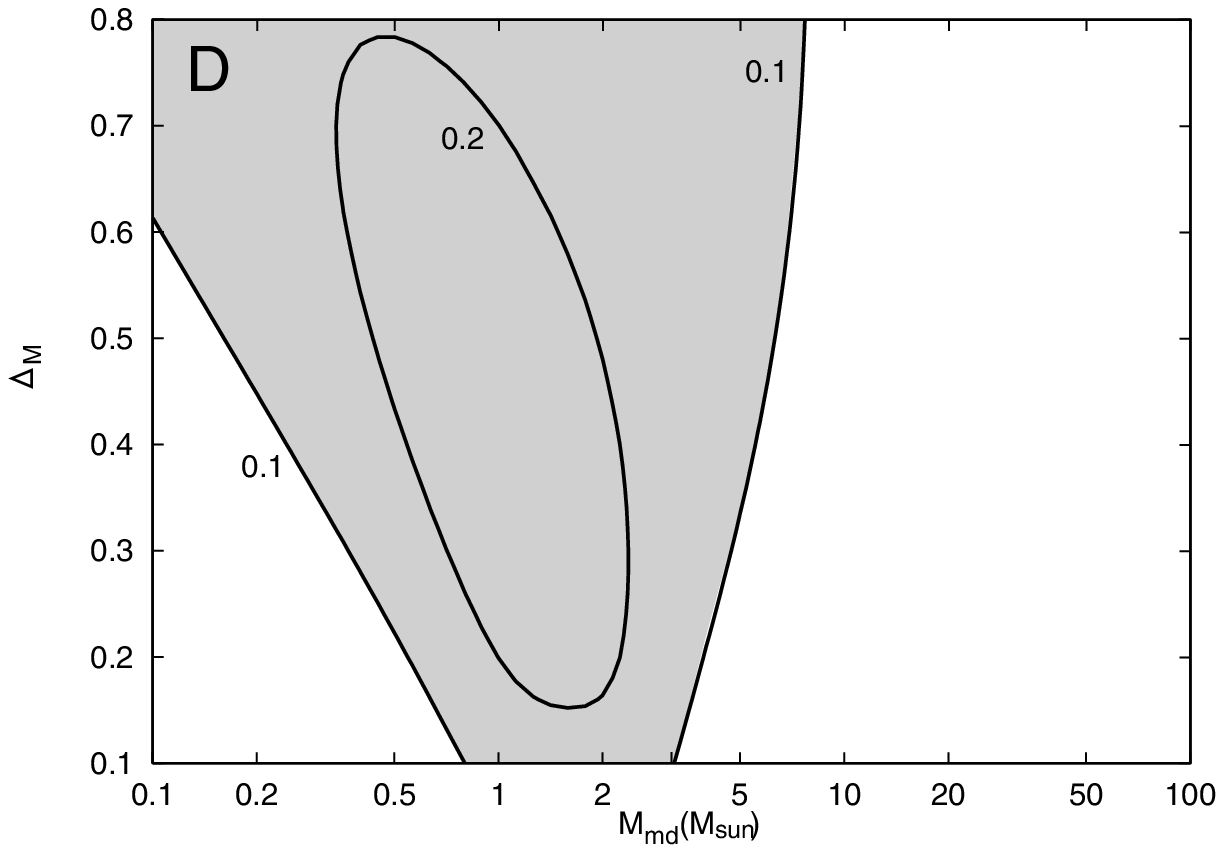}
\end{minipage}
\caption{
Dependence of the fraction of CEMP-$s$ stars on the medium mass, $\mmd$, and dispersion, $\dm$ of IMF of EMP population. 
Solid line denote the contour line of the fraction of CEMP stars among EMP sourvivors ($\psi_{{\rm CEMP\hbox{-}}s}(\mmd, \mmd) / \psi_{\rm{surv}}(\mmd, \dm)$). 
   Attached numerals designate the fractions.  
   Shaded area denote the parameter ranges for the IMFs that can give rise to the observed fraction of CEMP-$s$ stars. 
}
\label{fig:s}
\end{figure*}

\begin{figure*}
\epsscale{1.0}
\begin{minipage}{0.5\hsize}
\plotone{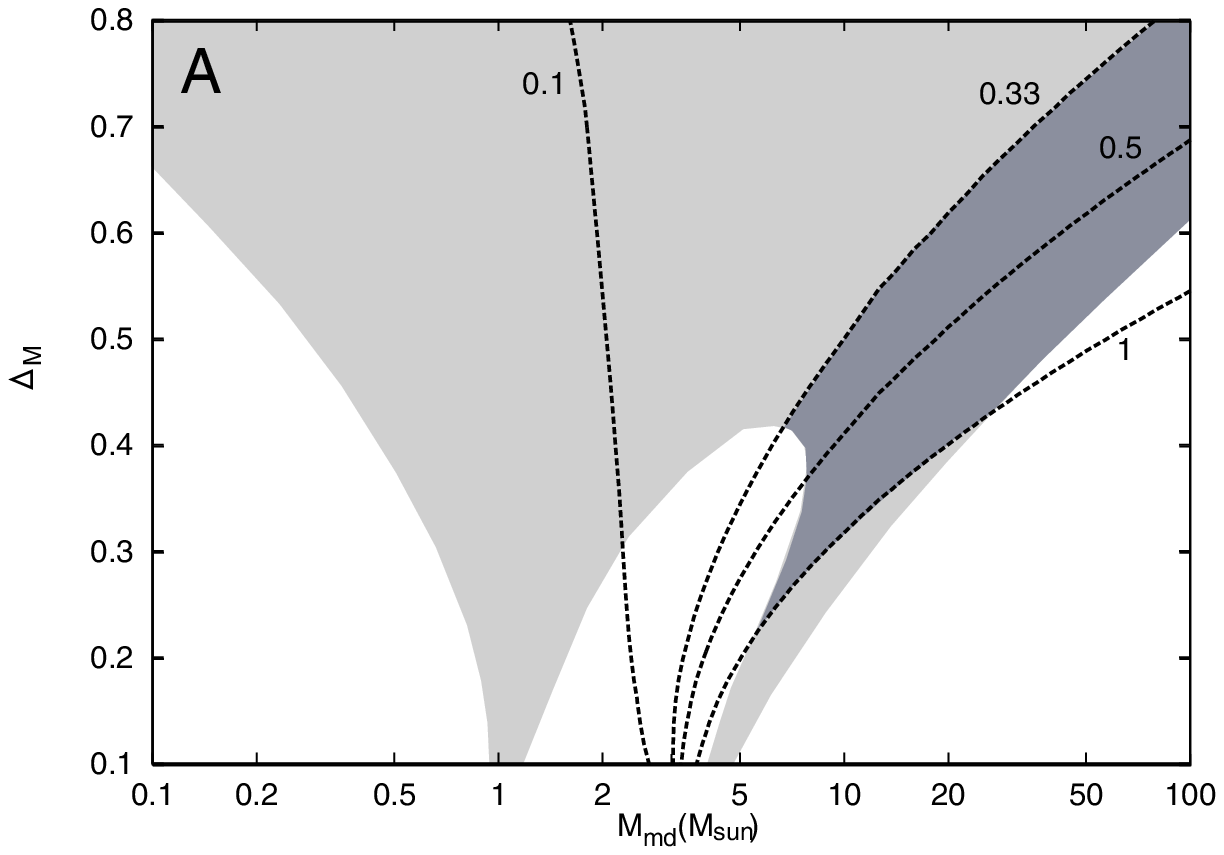}
\end{minipage} 
\begin{minipage}{0.5\hsize}
\plotone{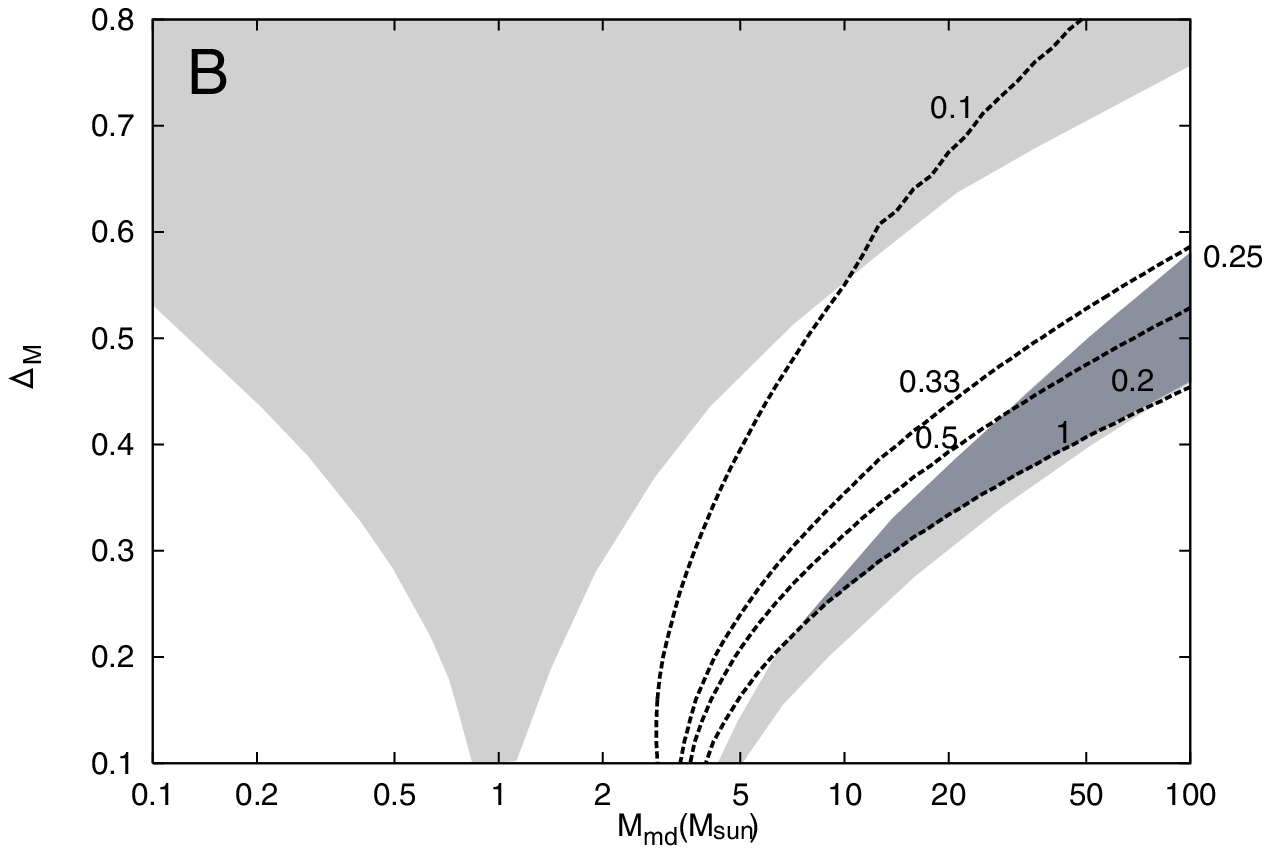}
\end{minipage} 
\begin{minipage}{0.5\hsize}
\plotone{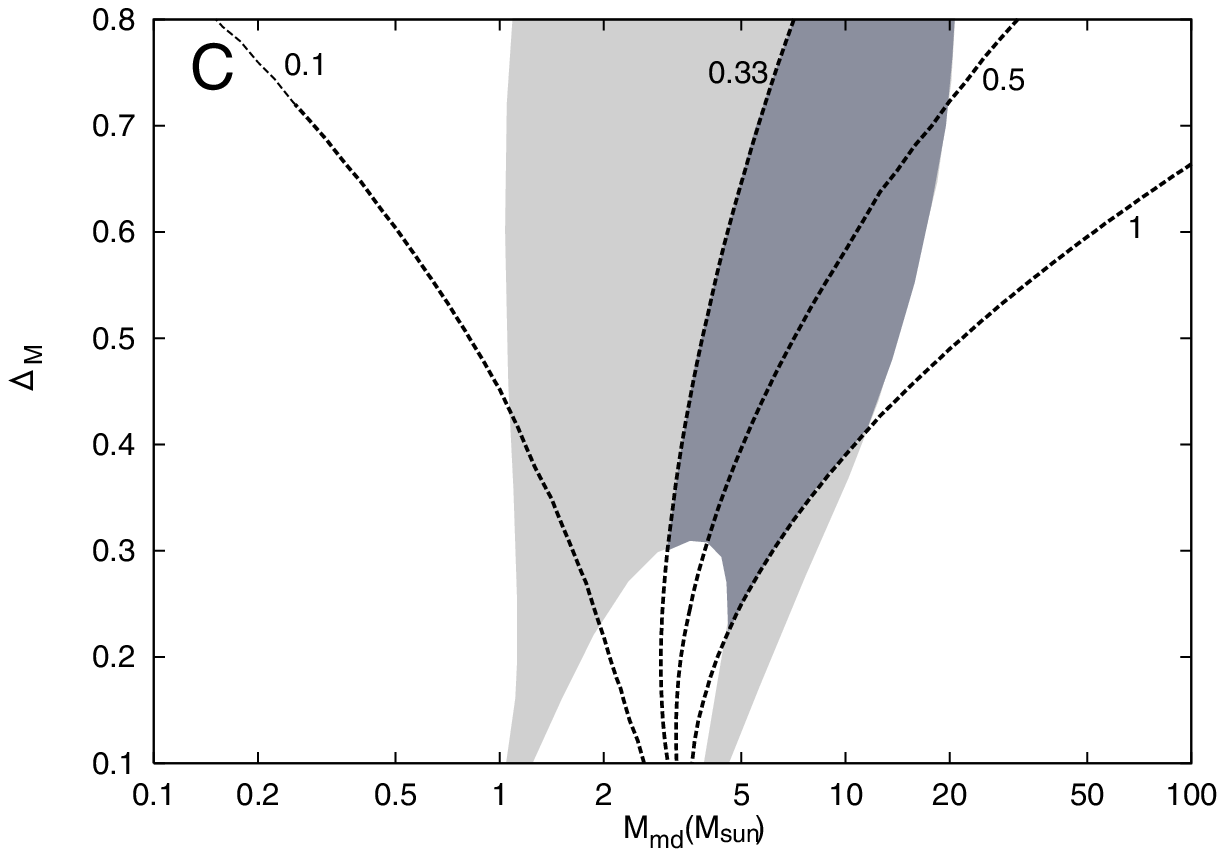}
\end{minipage}
\begin{minipage}{0.5\hsize}
\plotone{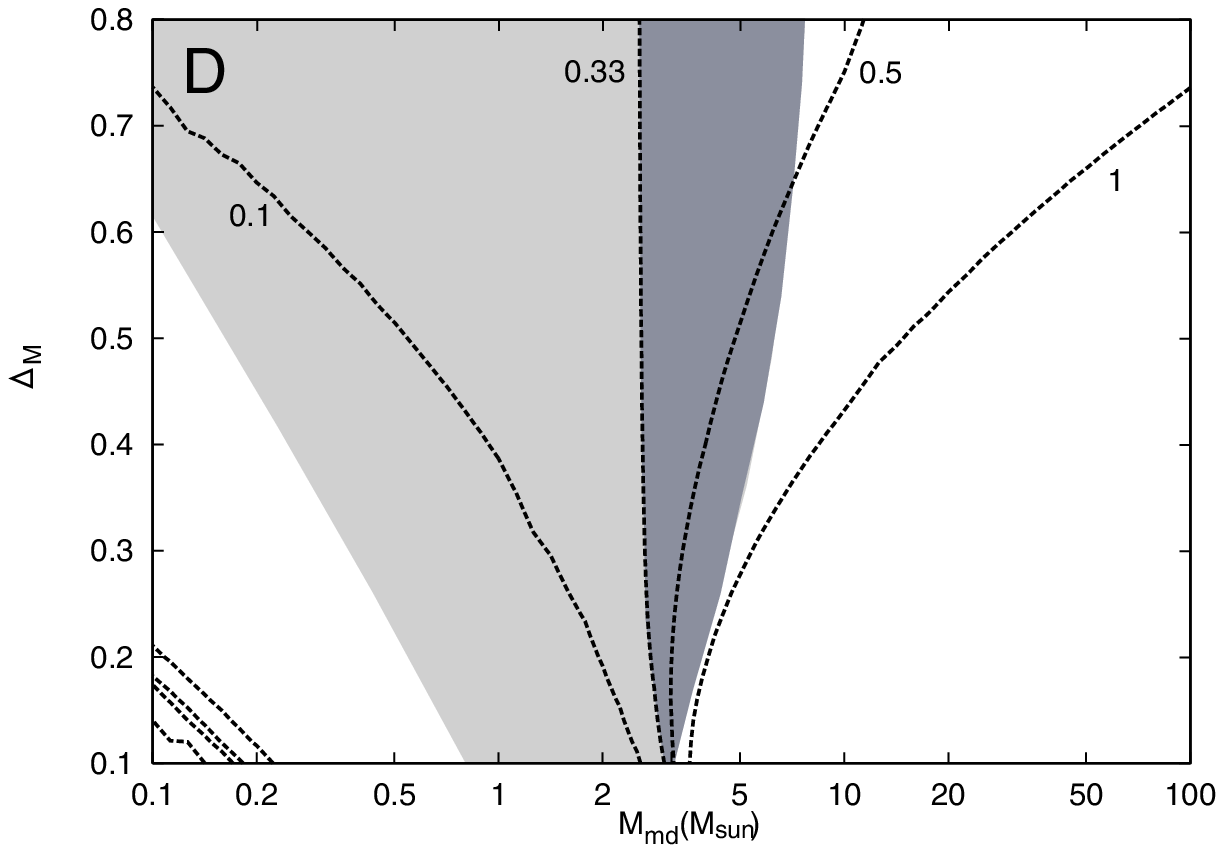}
\end{minipage}
\caption{
The contour line of the number ratio between \cemps\ stars and \cempnos\ stars ($\psi_{{\rm CEMP\hbox{-}no}s}(\mmd, \dm) / \psi_{{\rm CEMP\hbox{-}}s}(\mmd, \dm) $) on the $\mmd - \dm$ plane for cases A({\it Top left}), B({\it Top right}), C({\it Bottom left}) and D({\it Bottom right}), respectively. 
   Attached numerals designate the fractions. 
   Light shaded area denote the constraint from observed fraction of CEMP-$s$ stars as plotted in fig.\ref{fig:s}. 
   Dark shaded area denote ranges for the IMFs that fulfill the both constraints for the fraction of \cemps\ and the ratio between \cempnos\ and \cemps\ . 
}
\label{fig:nos}
\end{figure*}

For the four mass-ratio distributions, formulated in \S\ref{subsec-modelpara}, we can figure out, as the function of $\mmd$ and $\dm$, the portion of stars that survive to date ($M \le 0.8 \msun$), and then, the fractions of stars in these EMP survivors that evolved to \cemps\ and \cempnos\ stars according to the masses of primary stars and to the orbital separations.  
   Figure~\ref{fig:psi} compares the fractions of CEMP-$s$ stars in the EMP survivors [$\psi_{{\rm CEMP\hbox{-}}s} / \psi_{\rm surv}$] and the ratios of \cempnos\ to \cemps\ stars [$\psi_{{\rm CEMP\hbox{-}no}s} / \psi_{{\rm CEMP\hbox{-}}s}$], predicted with use of these four different mass-ratio distributions, as a function of medium mass $\mmd$ of IMFs with the dispersion of $\dm = 0.33$, taken to be same as the present-day IMF of Galactic spheroid component \citep{Chabrier03}.
   Figures~\ref{fig:s} and \ref{fig:nos} present the contour maps on the $\mmd$-$\dm$ diagram for the fractions of CEMP-$s$ stars and the ratio between the \cempnos\ and \cemps\ stars, respectively.  

Left top panels on these figures show the results for the flat mass-ratio distribution of Case~A, which reproduces the results obtained in Paper~I.  
   In Fig.~\ref{fig:psi}, the CEMP-$s$ fraction peaks at $\mmd = 4.8 \msun$, slightly above the upper mass limit of the primary stars for CEMP-$s$. 
   Note that when the secondary mass is specified, the mass distribution of primary stars peaks at mass smaller than $\mmd$ for this mass-ratio function [$\propto \xi (m_1) /m_1$, see Fig.~12 in Paper~I).  
   Two ranges of $\mmd$, $0.61-2.8 \msun$ and $ 7.6-15.3 \msun$ (light shaded parts) gives the IMFs compatible with the observations, separated by the overproduction of \cemps\ stars.  
   The relative frequency of \cempnos\ to \cemps\ stars is a steep increase function of $\mmd$, and excludes the lower range of $\mmd$ compatible with the \cemps\ fraction.  
   The IMFs with $\mmd = 4.8 -11.6 \msun$ (dark shaded part) gives compatible ratio with the observations 
   This range of $\mmd$ lies in the mass range of primary stars of \cempnos\ stars or even larger. 
   Accordingly, the intersection of the light and dark shaded parts designates the ranges ($\mmd = 7.6 -11.6 \msun$) that can explain the both statistics of CEMP stars, and hence, high-mass IMFs results for a dispersion $\Delta _M = 0.33$. 

On the $\mmd-\Delta_M$ diagram of Fig.\ref{fig:s}, the parameter space compatible with the observed \cemps\ fraction separates into two ranges for the dispersion smaller than $\dm \simeq 0.43$, converging to the narrow ranges around $M_{\rm md} \simeq 1$ and $4 \msun$, respectively, as $\dm$ decreases. 
   For larger dispersion, on the other hand, it merges into one part to cover wider range.  
   As for the ratio between the \cempnos\ and \cemps\ stars, Fig.~\ref{fig:nos} shows that the medium mass compatible with observed ratio increases with the dispersion to cover wider range, from $M_{\rm md} = 3.2 - 3.7 \msun$ at $\Delta_M = 0.1$ through $M_{\rm md} = 12.3 - 100 \msun$ at $\Delta_M = 0.54$.  
   Accordingly, for the IMFs that satisfy the both statistical constraints, the medium mass increases with the dispersion from $M_{\rm md} \ge 5.5 \msun$ for $\Delta_M = 0.22$ and beyond $\mmd = 100 \msun$ for $\dm > 0.62$.  

For the mass-ratio distribution function increasing with $q$ of Case~B (right top panel), the portion of binaries that have the secondary stars surviving to date decreases with the mass of primary stars in proportion to $(m_2 / m_1)^{-2}$, more steeply than in proportion to $(m_2 /m_1)^{-1}$ for a flat mass-ratio distribution in Case~A.  
   Since the average mass of the primary stars is smaller for a given EMP star, therefore, the fraction of CEMP-$s$ stars is larger for a given $M_{\rm md}$, and the peak shifts to larger $M_{\rm md}$, as compared with Case~A.  
   In Fig.~\ref{fig:psi}, the $M_{\rm md}$ of IMFs compatible with the observed fraction of \cemps\ stars separates into two ranges, as in Case~A, but the in-between gap is larger; 
   the higher mass range shifts upward in mass to greater extent ($\mmd = 14.5 - 28\msun$) than the smaller mass range shifts downward ($\mmd = 0.38 - 2.4 \msun$).  
   This also causes a smaller ratio of \cempnos\ to \cemps\ stars for a given $\mmd$, and hence, the IMFs compatible with the observed ratio shift to a larger mass of $M_{\rm md} = 8.9 - 21.6 \msun$, as compared with that for Case~A.  
   As a result, the IMFs consistent with the both statistics of CEMP stars turns out to be higher mass by a factor of $\sim 2$ than for Case~A ($\mmd = 14.5 - 21.6 \msun$ for $\dm = 0.33$). 
   In the Fig.~\ref{fig:s}, we see that the range of $\mmd$, compatible with the observed fractions of \cemps\ star (shaded area), separates into two and the higher range shifts to larger mass for a given $\dm$.  
   Similarly, in the Fig.~\ref{fig:nos}, the observed ratio of the \cempnos\ to \cemps\ stars also demands larger $\mmd$, and the range of $\mmd$ of IMFs compatible with the observation increases rapidly with $\dm$ to exceed $100 \msun$ for $\Delta_M \ge 0.59$. 
   In order to satisfy the both conditions of CEMP star observations, the IMFs fall in the range of higher medium mass and in a rather narrow range of dispersion, lying in the parameter space of $\mmd > 7.1 \msun$, larger by a factor of $\sim 1.3$ than for Case~A, and $\dm > 0.22$ and of $\dm = 0.45 - 0.58$ for $\mmd=100 \msun$.  

For a mass ratio function decreasing with $q$ of Case~C,\red{ we see the opposite tendency of Case~B (bottom left panels of Fig.~\ref{fig:psi} - \ref{fig:nos}). }
   The portion of EMP binaries whose low-mass members survive to date depends only weakly on the primary mass ($\propto \log m_1$) so that the fraction of CEMP-$s$ stars reduces because of larger contributions from the binaries with more massive primaries.  
   As seen in Fig.~\ref{fig:psi}, the fraction of \cemps\ stars in the total EMP survivors is well below the upper bound of the observations, and hence, the $\mmd$ compatible with the observations merges into one narrower range of $\mmd = 1.1 - 15.6 \msun$ for $\dm = 0.33$.  
   The observed ratio of \cempnos\ to \cemps\ stars can be reproduced also by the IMFs with a smaller $\mmd$ by a factor of $\sim 2$ than in Case~A ($\mmd = 3.2 - 7.5 \msun$). 
   Accordingly, the $\mmd$ for the IMFs consistent with the both CEMP star statistics are smaller by a factor of $1.5 -2.4$ than for Case~A (the mass range $\mmd = 3.2-7.5 \msun$ for $\dm =0.33$).  
   In the Fig.~\ref{fig:s}, the range of $\mmd$ for the IMFs, compatible with the observed \cemps\ fraction varies only little with $\dm$, and is restricted in the range between $\mmd = 1.1 - 23 \msun$, though it separates into two for small $\dm < 0.31$.  
   As shown in Fig.~\ref{fig:nos} the dependence of the ratio of \cempnos\ to \cemps\ stars on $\dm$ is also weaker than for Case~A. 
   Consequently, the IMFs can reproduce the both CEMP star statistics with the mass as small as $\mmd = 3.3 \msun$, smaller by a factor of $\sim 0.6$ than for Case~A, but differently from the above two cases, an upper bound is placed at $\mmd =23 \msun$, regardless of the dispersion with a lower bound of $\dm > 0.21$.  

Bottom right panels depict the results for the ``independent" coupling of Case~D. 
   For this case, the number of EMP survivors produced per binary is independent of the mass, $m_1$, of primary stars, while the binary frequency itself increases with $m_1$.  
   The former is similarly to Case~C, and then, the production of EMP survivors from the binaries with massive primary poses a severe constraint on the high-mass side of IMFs.  
   On the other hand, the latter favors the production of \cemps\ stars as compared with the low-mass binaries of $m_1 \le 0.8 \msun$. 
   The both shift the IMFs, compatible with the observed fraction of \cemps\ stars, to smaller $\mmd$.   
   In addition, the single stars, born in the same number of binaries, contribute to significant fraction of EMP survivors, increasing from $1/3$ up to $1/2$ for smaller $\mmd$ for $\mmd < 0.8 \msun$ since the low-mass binaries are counted as one object.    
   As a result, the maximum fraction of \cemps\ stars remains below the upper limit of the observed range, which makes the $\mmd$ for the IMFs that can reproduce the observation lie in a single range within a relatively small upper bound.  
   The observed ratio of \cempnos\ to \cemps\ stars demands also lower-mass IMFs, as for Case~C.  
   Accordingly, the IMFs that can reproduce the both CEMP star statistics fall in the narrowest range of $\mmd = 2.5-7.0 \msun$ with rather small upper mass limit, almost irrespective of the dispersion, on the $\mmd-\dm$ diagram in Fig.\ref{fig:nos}.  
   The \cemps\ star faction remains smaller than $\sim 20 \%$ because of the contribution of the stars born as single. 
   
In conclusion, the statistics of CEMP stars demand the IMFs for the EMP population, peaking at the intermediate-mass stars or the massive stars, by far higher mass than those of Pop.~I and II stars, irrespectively of the assumed mass-ratio distribution.  
   The presence of \cempnos\ stars in a significant number of the \cemps\ stars excludes the IMFs of small mass.  
   The derived mass range varies by a factor of $\sim 2$, from the highest $\mmd > 7.1 \msun$ for the mass-ratio distribution of increase function of the mass ratio (Case~B) to the lowest $2.5 < \mmd / \msun < 7$ for the mass-ratio distribution of "independent'' coupling (Case~D).  
   This tendency is explained in terms of the difference in the averaged mass of the primary companion of the EMP survivors;  
   if the contributions to the EMP survivors decrease rapidly with the mass of primaries, a relatively higher-mass IMFs result without an upper mass limit imposed, while if the contribution to the EMP survivors are weakly dependent or independent on the primary masses, an upper limit is set with the relatively smaller-mass on the IMFs.    

\subsection{Constraints from Galactic Chemical Evolution}\label{CEsec}
   \red{In this section, we discuss that a}dditional constraints can be derived \red{from the relationship between} the total number of EMP survivors and the iron yields from the EMP population.

\begin{figure*}
\epsscale{1.0}
\begin{minipage}{0.5\hsize}
\plotone{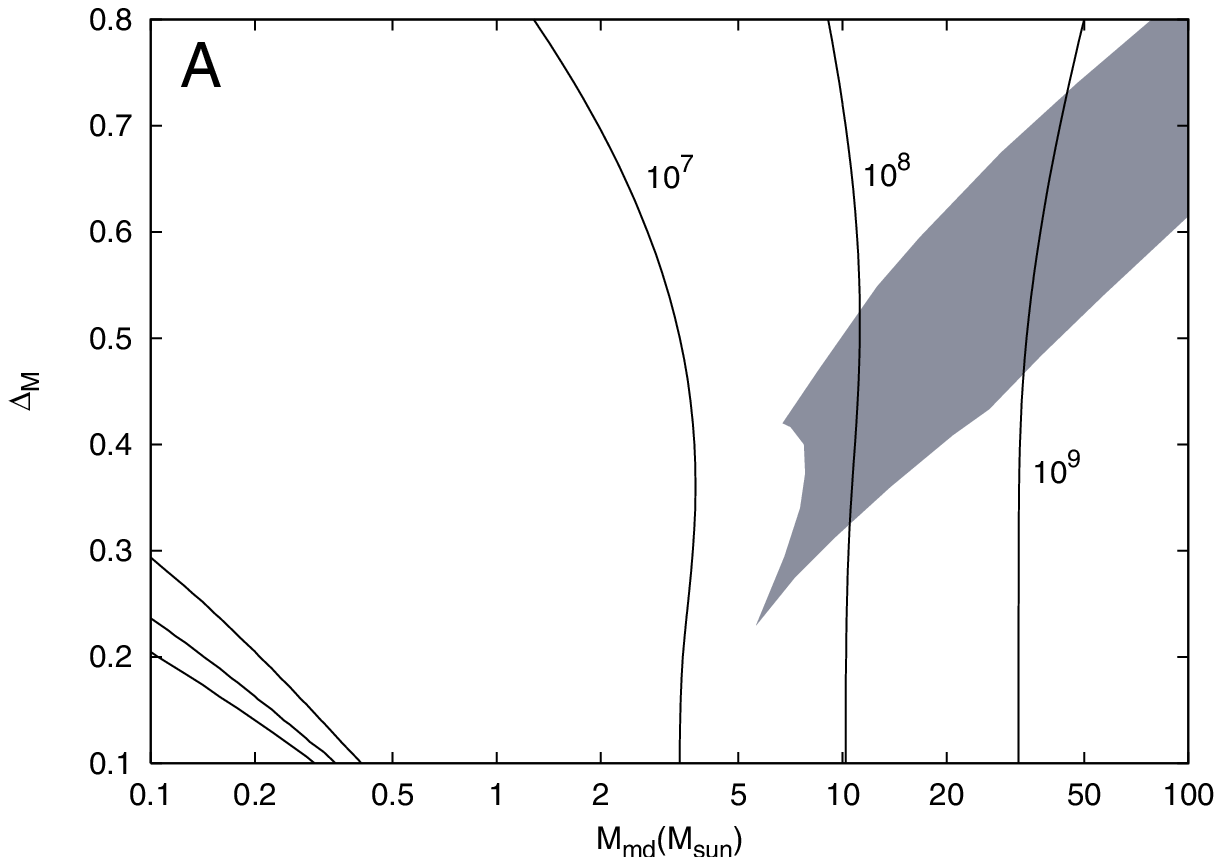}
\end{minipage} 
\begin{minipage}{0.5\hsize}
\plotone{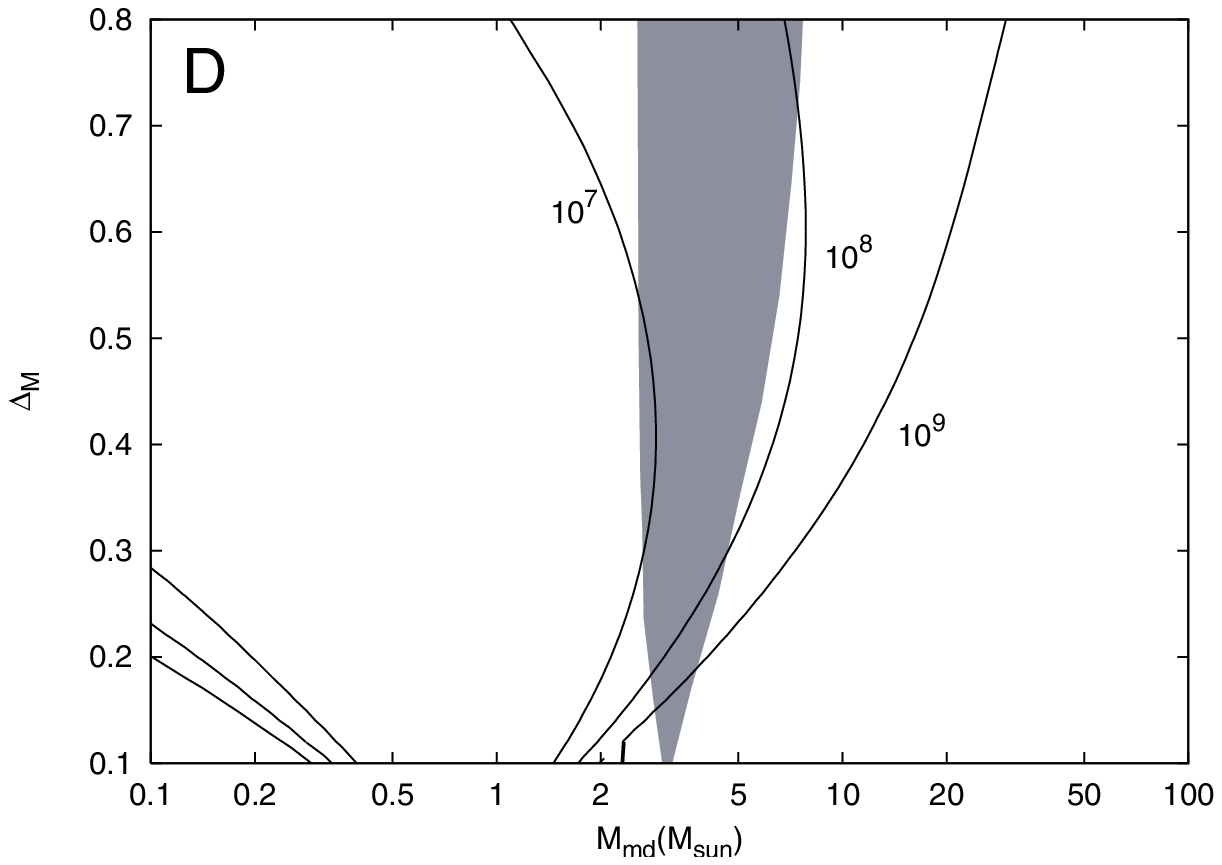}
\end{minipage} 
\caption{
The contour of the total mass of EMP population stars, $M_{\rm EMP} = 10^7, 10^8$ and $10^9\msun$ on the diagram of parameters (medium mass $\mmd$ and dispersion $\dm$) in the log-normal form for the flat distribution (Case~A: left pannel) of the mass ratio and independent counling (Case~D: right pannel). 
  Shaded area denotes the parameter range consistent with the statistics of CEMP stars.   
}
\label{fig:Memp}
\end{figure*} 

\begin{figure*}
\epsscale{1.0}
\begin{minipage}{0.5\hsize}
\plotone{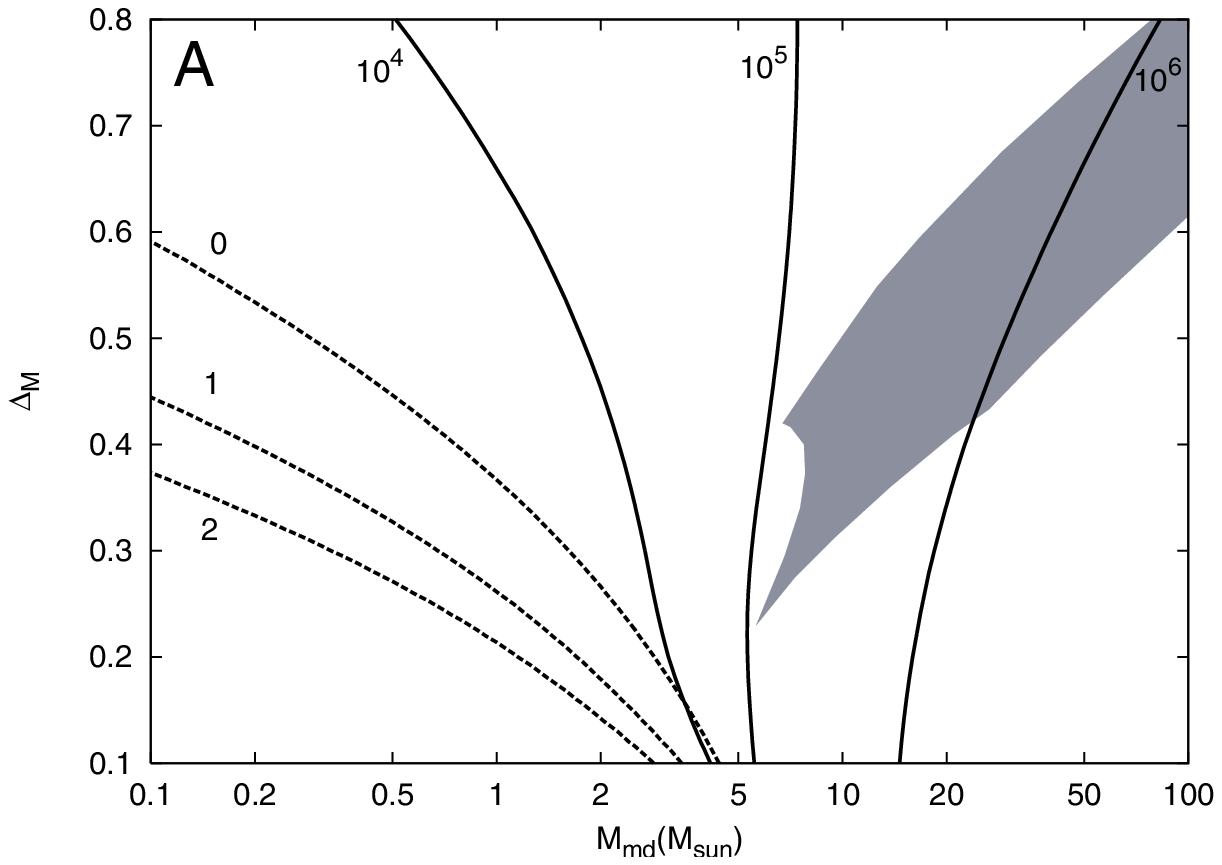}
\end{minipage} 
\begin{minipage}{0.5\hsize}
\plotone{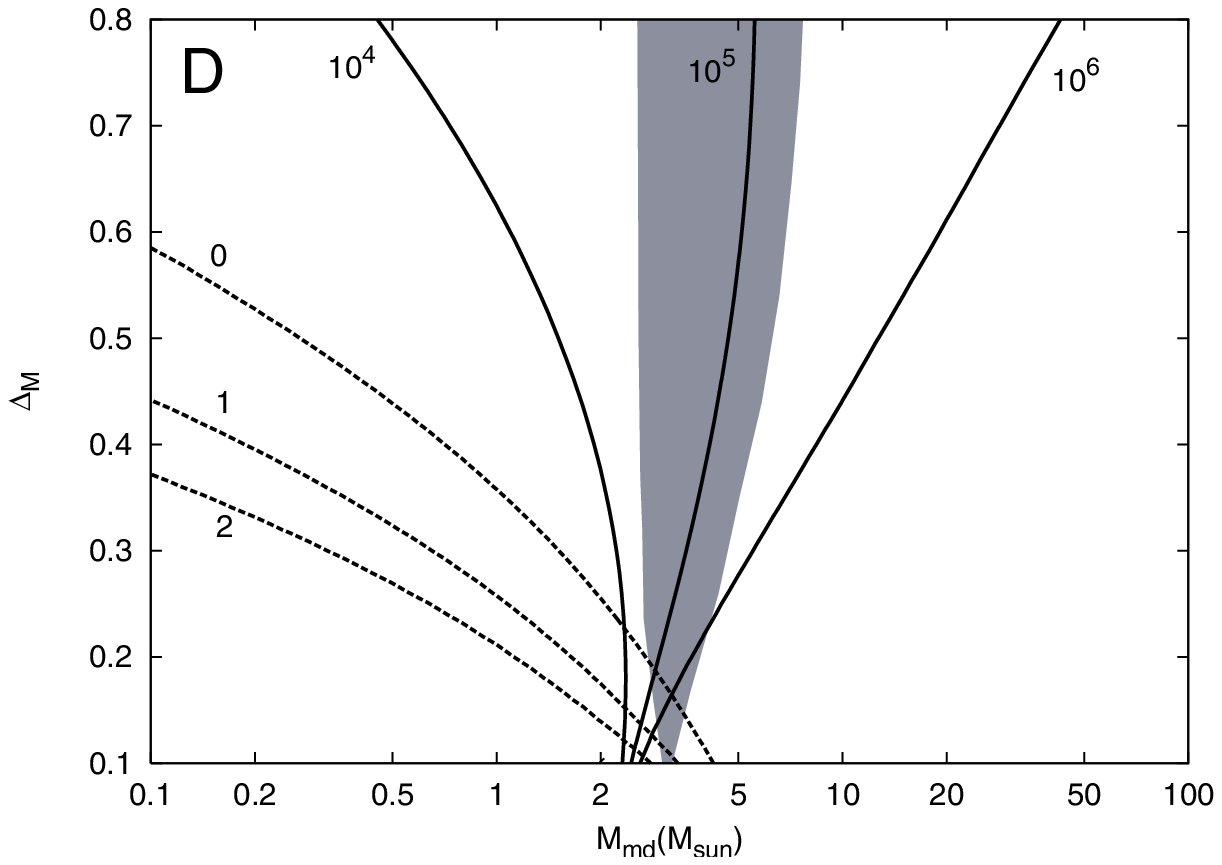}
\end{minipage} 
\caption{Constraints on the IMF of EMP population, derived from the number of EMP survivors and the iron production by the stars of EMP population on the $\mmd - \dm$ plane. 
  Solid lines denote, the loci of IMFs which can produce the amounts of metal production, $M_{\rm Fe, EMP} = 10^4, 10^5, $ and $10^6 \msun$.  
  Dashed lines denote the contour of carbon production by AGB stars to the iron production, $\abra{C}{Fe} = 2$, 1, and 0.  
}
\label{CEfig}
\end{figure*}  

\blue{
Figure~\ref{fig:Memp} shows the contour maps of the total stellar mass, $M_{\rm EMP}$, of EMP population, \red{in eq.~(\ref{eq:EMP-mass}),} necessary to leave the observed number of EMP survivors on the $\mmd$-$\dm$ diagram. 
   The total stellar mass increases for higher mass IMFs to produce the given number of low-mass survivors.  
   For the flat mass-ratio distribution (Case A; left panel), the total stellar mass of EMP population is essentially determined by $\mmd$ in proportion to $\mmd^2$ and only weakly dependent on $\dm$ for large $\mmd$, since almost all EMP survivors formed as secondary (i.e. second term of eq.~[\ref{eq:frac-giants}] is dominant) for IMF of $\log (\mmd/0.8 \msun) \gg \dm$. }
   Figure~\ref{CEfig} shows the contours of the total iron mass, $M_{\rm Fe, EMP}$, produced by the massive stars of EMP population, \red{in eq.~(\ref{eq:EMP-iron}). 
   The iron production also increases with larger $\mmd$ but is more sensitive to $\dm$ since the dependence of the supernovae fraction differs across the border of $\mmd \simeq M_{up}$;  
   the amount of produced iron increases (or decreases) with $\dm$ for given $\mmd < M_{up}$ (or $\mmd > M_{up}$), little dependent on $\dm$ for $\mmd \simeq M_{up}$.   }

\blue{For the "independent'' coupling (Case~D; left panel), the fractions of giant EMP survivor and supernova among EMP population stars are given by, 
\begin{eqnarray}
& f_{\rm G} & = (1+f_b) \xi (0.8) \Delta M_G \\
& f_{\rm SN} & = (1+f_b) \int_{M_{up}} dm_1  \xi (m_1)
\end{eqnarray}} 
   The total mass and the amount of iron production of EMP population are sensitive both to $\mmd$ and $\dm$, especially for small $\dm$ and large $\mmd$ in contrast to with the other cases. 
   For small $\dm$, therefore, the fraction of low-mass stars varies greatly with $\mmd$, and the both contours of $M_{\rm EMP}$ and $M_{\rm Fe}( M_{\rm EMP})$ converge to $\mmd \simeq 2-3 \msun$.  
   As $\dm$ increases, the differences from Case~A diminish since the IMFs tend to extend into the low-mass stars, and in particular, for $\dm \gtrsim 0.4$ and $\mmd \lesssim 3 \msun$, the contours in the both panels resemble each other to run through the similar parameter spaces.  
   
From the comparison with the total amount of iron $M_ {\rm Fe, EMP}$, necessary for the chemical evolution, in this diagram, the parameter space where $M_{\rm Fe, EMP} (\mmd, \dm) \gg M_ {\rm Fe, halo} = 10^{5.5} \msun$ is excluded by the overproduction of iron or by the underproduction of EMP survivors. 
   For the parameter space where $ M_{\rm Fe, EMP} (\mmd, \dm) \ll M_ {\rm Fe, halo}$, on the other hand, the stars of EMP population can leave the number of EMP survivors currently observed but are short of iron production, so that the chemical evolution demands other sources of iron production without producing the low-mass stars that survive to date. 
   For a flat mass-ratio distribution, the IMFs that can satisfy the condition of iron production coincide the IMFs, derived above from the statistics of CEMP stars (shaded area) in the parameter range of $\mmd \simeq 10 - 16 \msun$ and $\dm \simeq 0.3 - 0.6$.  
   For the ``independent" coupling, the parameter range of IMFs that satisfy the condition of iron production also overlap the shaded area of parameter range, derived above from the statistics of CEMP stars, but with the mass $\mmd \simeq 3.5-5.1 \msun$, slightly smaller than for Case~A and only for a small dispersion of $\dm \simeq 0.23-0.35$.  
   For larger $\dm$, even the highest-mass IMFs of $\mmd =7.5 \msun$ result to be slightly short of, or marginally sufficient at the most, iron production.  

For the two other mass-ratio distributions of $n\propto q$ (Case~B) and $n \propto 1/q$ (Case~C), the iron production, $M_{\rm Fe, EMP}$, with a given IMF results to be larger or smaller than for Case~A because of the difference in the number of massive stars exploded as supernova per low-mass survivor (e.g., by a factors of 1.38 and 0.47, respectively, per a star of $m=0.8 \msun$ and the IMF of $\mmd=10\msun$ and $\dm =0.4$).  
   The iron production then demands smaller-mass (higher-mass) IMFs for Case~B (Case~C) as compared with Case~A, the shift of IMFs in an opposite direction, discussed from the statistics of CEMP stars. 
   Accordingly, for these two extreme cases, the parameter ranges for the IMFs derived from the statistics of CEMP stars and the iron production are marginally overlapped (Case~C) or are dislocated with a narrow gap (Case~B), although a definite conclusion waits for future observations, in particular, to improve the estimate of total numbers of EMP stars (see Appendix).  

The relative production rate of carbon to iron may also impose additional constraint since the intermediate-mass stars enrich intergalactic matter with carbon through the mass loss on the AGB, as discussed by \citet{Abia01}. 
    In particular, when $\dm$ is small and $\mmd$ is in the range of intermediate- and low-masses, the intermediate-mass stars much surpass the massive stars in number and eject more carbon than the latter eject iron.  
   We compute the amount of carbon ejected by AGB stars by taking the carbon abundance in the wind ejecta of AGB stars at $\abra{C}{H}=0$, and the remnant mass at $1 \msun$.  
   Contours of $\abra{C}{Fe} = 2, 1, 0$ are plotted in the figure (dashed lines), for which only the carbon from the AGB stars are taken into account.  
   The overabundance of carbon excludes the IMFs with low dispersion and low medium mass;  
   it excludes the parameter space in the range of $\dm < 0.2$, derived by the CEMP star statistics for Case~D, but has nothing to do with the high-mass IMFs derived for Case~A.  

We demonstrate that the IMFs, derived from the observed properties of CEMP stars, have the parameter ranges that can explain the chemical evolution and the production of low-mass stars, consistent with the observations, both for the flat mass-ratio distribution and for the ``independent" coupling.  
   In Appendix we will discuss the converse to demonstrate that the argument based on the total number of EMP survivors and the total iron production can potentially provides more stringent constraint on IMFs, independent of the argument based on the CEMP star statistics. 

\blue{Relative abundances of other elements may also be affected by the IMF. } 
\red{   Theoretical study of supernova nucleosynthesis suggests the peculiarities and variations of yields for the metal-free and extremely metal-poor stars \citep{Woosley95, Umeda02, Heger02,Tominaga07,Heger08}.
    The supernova yields are, however, sensitive to the assumption of model parameters such as the explosion energy and the treatment of non-axisymmetric effects, and currently subject to the large uncertainties. }
    In this paper, therefore, we \red{are concerned with the iron yields as an indicator of the chemical evolution, and defer detail study about the abundance pattern of various elements in future works. 
    As for the iron yield, recently, type Ia supernovae with short delay time and their contribution of iron production are discussed by some authors \citep[e.g.][]{Scannapieco05}, although the evolutionary scenario is not yet clear.  
   The iron reduction is suggested to be several times larger than by type II supernovae, and yet, will hardly affect our results since the ratio, $f_{\rm SN} / f_{\rm g}$ depends strongly on the IMF. } 

\subsection{Distinctive Features of EMP Survivors}

The different assumptions on the mass-ratio distributions admit the parameter ranges of high-mass IMFs that can reproduce the statistics of CEMP stars and the chemical evolution, consistent with the existent observations.   
  The predicted mass ranges differ by a factor of 2 or more between $\mmd \simeq 5-20 \msun$. 
  Although hardly distinguishable from the observations discussed so far, they surely make the differences in the properties of EMP survivors. 
   We discuss the imprints that the mass-ratio distributions have left on the current EMP survivors and investigate the possibility of discriminating the mass coupling of binary systems in the EMP population, especially for the two distinct distributions of the flat mass-ratio distribution and the ``independent'' coupling.  

\begin{figure}
\epsscale{1.0} \plotone{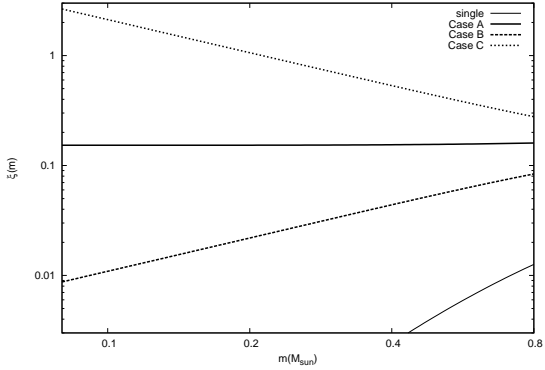}
\caption{The mass function of EMP survivors under the different assumptions on the coupling mass distribution of binaries. 
  Here the parameters of IMFs are taken to be $\mmd = 10 \msun$ and $\dm = 0.4 $ (Cases~A-C). }
\label{fig:survivors}
\end{figure}

Firstly, an obvious difference is the mass distribution function of EMP survivors.  
   For a given IMF, $\xi (m)$, the mass distribution, $\xi_{\rm EMP\hbox{-}surv} (m)$, of EMP survivors is given by;
 \begin{eqnarray}
\xi_{\rm EMP\hbox{-}surv} (m) &=& (1-f_b) \xi (m) + f_b \xi (m) \int^m n(m_2/m) /m dm_2 \nonumber \\
&+& f_b \int_{0.8 \msun}\xi (m_1) n(m/m_1) /m_1 d m_1.
 \end{eqnarray}
   Here a low-mass binary, whose components are both less massive than $0.8 \msun$, is counted as one object with the primary star.
   Figure~\ref{fig:survivors} shows the mass distributions of EMP survivors ($m \le 0.8 \msun)$ for different assumptions of mass-ratio distributions Cases~A-C.  
   For these mass-ratio functions, the mass distribution of EMP survivors is nearly proportional to the mass-ratio distribution $n(q)$ because almost all of them come from the secondary stars;  
   the contribution from the primary components are denoted by thin solid line, and the same contribution comes from the stars born as single. 
   For the ``independent" coupling, in contrast, the $\xi_{\rm EMP\hbox{-}surv} (m)$, has the same form as the IMF and the number of EMP survivors decreases rapidly as the stellar mass decreases.

Secondly, the fraction of double-lined binary and the contribution of stars born as single among EMP survivors may differ according to the mass-ratio distribution.  
\blue{The EMP survivors born as binaries are divided into three categories according to the mass of the primary stars:
   (1) the low-mass binaries with the primary of mass $m_1 \le 0.8 \msun$, 
   (2) the white-dwarf binaries of primary stars of mass between $0.8 \msun < m_1 \le M_{up}$, and 
   (3) the supernova binaries of primary stars of mass $m_1 > M_{up}$. 
   The fraction of low-mass binaries with the primary stars of mass $ m \le 0.8 \msun$ in the EMP survivors of mass between $m$ to $m + dm$ is given by
\begin{equation}  
  \varphi_{\rm surv, LMB} (m) = f_b [\xi (m) /m] \int^m n(m_2/m) dm_2 /\xi_{\rm EMP\hbox{-}surv} (m).
\end{equation}  
   They can be detected as double-lined binary. }
   For the flat mass-ratio distribution, this gives a significant fraction of $ \varphi_{\rm surv, LMB} (0.8 \msun) = 7.3 \%$ for $\mmd= 10 \msun$ and $\dm =0.4$, and increases with $\dm$ to $16 \% $ for $\dm = 0.5$ and with decreasing $\mmd$ to 18\% for $\mmd= 5 \msun$, respectively.  
   \blue{We note that these values depend weakly on $f_b$ since most of the EMP survivors are from the binaries. }
   The number of low-mass binary decreases rapidly for smaller masses while the number of EMP survivors, formed as the low-mass members of white dwarf binaries or supernova binaries, remains constant.  
   
For the ``independent" coupling, the fraction of low-mass binaries in the EMP survivors reduces to;
\begin{eqnarray}
\varphi_{\rm surv, LMB} (m) = 2f_b \int^{m } \xi (m_2) dm_2 \Bigg/ \Bigg[(1+f_b) \nonumber \\
- 2f_b \int_m^{0.8 \msun} \xi (m_1) d m_1\Bigg], 
\end{eqnarray}
   which gives a much smaller fraction of $ \varphi_{\rm surv,LMB} (0.8 \msun) = 1.6 \%$ for $\mmd= 5 \msun$ and $\dm = 0.4$ as compared with the flat mass-ratio distribution. 
  The fraction may increase for smaller medium mass, to 5.5\% at $\mmd = 3\msun$, and for larger dispersion, to 3.9 \% and 9.5\% at $\dm=0.5$ and 0.7, respectively, although these may cause underproduction of iron, in particular for smaller $\mmd$, as seen from Fig.~\ref{CEfig} (bottom panel).  
  In this case, the proportion of the EMP survivors, born as single stars, is fairly large as given by 
\begin{equation}
   \varphi_{\rm surv, sing} (m) \simeq (1-f_b) \Bigg/ \left[(1+f_b) -2f_b \int_m^{0.8 \msun} \xi (m_1) d m_1\right].   
\end{equation}
  Consequently, nearly one third of EMP stars were born as single stars, for $f_b=0.5$, which is much larger fraction than in the case of the flat mass-ratio distribution.  

Thirdly, the fraction, $\varphi_{\rm surv, SNB}$, of supernova binaries with the primary stars of mass $m_1 > M_{up}$ also differs between the two mass-ratio distributions.  
   For the flat mass-ratio distribution, almost all of the EMP survivors belong, or have been belonged, to the binary systems, and the fraction is given by  
\begin{equation}  
  \varphi_{\rm surv, SNB} (m) = f_b \int_{M_{up}} n(n/m_1) \xi (m_1) / m_1 d m_1 / \xi_{\rm EMP\hbox{-}surv} (m),
\end{equation}
   and amounts to $\sim 50\%$.  
   For the `` independent" coupling, on the other hand, one third of EMP survivors are single stars from their birth, and the percentage of supernovae binaries is relatively small, as given by
\begin{eqnarray}  
  \varphi_{\rm surv, SNB} (m) &=& 2f_b \int_{M_{up}} \xi (m_1) d m_1 \Bigg/ \Bigg[ (1+f_b) \nonumber \\
&&-2f_b \int_m^{0.8 \msun} \xi (m_1) d m_1 \Bigg],
\end{eqnarray}
  and turns out to be $\sim 20 \%$. 
   The EMP survivors from the supernova binaries have experienced a supernova explosion of the erstwhile primary stars at close distances and are thought to suffer from some abundance anomalies, affected by supernova ejecta.  
   Accordingly, these stars, in particular from the binaries of sufficiently small separations, may be discriminated by a large enhancement of elements, characteristic to the supernova yields. 

These differences in the properties of remnant EMP survivors may potentially serve as tools to inquire into the nature of EMP binaries and to distinguish the mass-ratio distributions.  
   Among the EMP stars, several double-lined spectroscopic binaries are reported in the literature. 
   If we restricted to the metallicity range of $\feoh <-3$, for which the observations with high-resolution spectroscopy may be regarded as unbiased, there are two stars CS22876-032 \citep[$\feoh \simeq -3.6$, $V=12.84$, $P=424.7$ d, and $m_2/m_1 \simeq 0.89$;][] {Thorburn93,Norris00,Gonzalez08} and CS 22873-139 \citep[$\feoh \simeq -3.4$, $V=13.8$, $P=19.165$ d, $m_2/m_1 \simeq 0.92$;][] {Preston94,Preston00,Spite00} with the detailed analyses and one star HE 1353-2735 \citep[$\feoh \simeq -3.2$, $V=14.7$;][]{Depagne00} without the binary parameter.  
   So far 39 dwarf stars of $\feoh < -3$ are confirmed by the high-resolution spectroscopy (we define the dwarf as $\log g [\hbox{ cm s}^{-2}] \ge 3.5$), and hence, the fraction of low-mass binaries, composed of two unevolved EMP stars, turns out to be $\gtrsim 3 /39 \simeq 7.7 \%$. 
\blue{It seems to be compatible with the estimated fraction $\varphi_{\rm surv, LMB} (0.8 \msun) = 7.3\%$ for the frat mass-ratio distribution but to be a little larger, or marginal, for the ``independent" coupling.  
   We note, however, that only with two samples, the above fraction may be subject to significant observational selection effects.   
   These stars have to be concentrated near to the upper-end of main-sequence since they are found among the candidates, selected from the flux limited surveys, and the mass ratio has to be sufficiently large for the lines of two components to be observed. 
\red{   The actual fraction has to be larger than observed if} we take into account the detection probability due to the orbital phase and to the inclination angle, and the rather narrow range of mass-ratios for the observed double-lined binaries. 
   On the contrary, the larger survey volume by a factor up to $2^{3/2}$ for the double-lined binaries due to the sum of luminosities may reduce the actual fraction.  
   More observations for the main-sequence EMP binaries and the bias corrections are necessary to discriminate the mass ratio distributions. }

It may be more straightforward to compare our results with the mass distribution function of EMP survivors.  
  From the existent observations, however, it is rather hard to determine since the observed dwarfs are mostly concentrated near to the upper end of main sequence. 
   An exception is a carbon dwarf G77-61 of $\feoh = 4.03$ \citep{Plez05} whose mass is inferred at $0.3 -0.5 \msun$, but it was found among the proper-motion-parallax stars \citep{Dahn77}, not from the surveys. 
   We have to wait for the larger-scaled surveys in near future to reveal the distribution of EMP survivors of low masses.  
   As for the supernovae binaries, they are expected to be related to the large star-to-star variations in the surface elemental abundances, in particular, with those of r-process elements, ranging more than by two orders of magnitude. 
   It is necessary, however, to understand the nature of interactions between the supernova ejecta colliding at very high velocity and the near-by low-mass stars before the meaningful conclusions can be drawn from the observations.  
  
\section{Metallicity Distribution Function of EMP Stars }

We have shown that the high-mass IMFs with the binary provide a reasonable explanation of the observed properties of EMP stars in the Galactic halo, revealed by the recent large-scaled HK and HES surveys. 
   In this section we discuss the consequence of derived IMF on the metal enrichment history of Galactic halo up to $\feoh=-2.5$ to study their relevance to the metallicity distribution function (MDF), observed for the EMP stars.  
\red{   As seen above, the derived IMFs using the different assumptions on the mass-ratio distribution function are hardly distinguishable by the current observations.  
   In the following, therefore, we assume the IMF with the medium mass $\mmd = 10\msun$ and the dispersion $\dm = 0.4$ in the lognormal form, which is at least compatible with any of the assumed mass-ratio distributions. } 

\subsection{Simple Model of Chemical Evolution}

Under the assumption that matter ejected from supernovae spreads homogeneously and is recycled instantaneously, the iron abundance, $X_{\rm Fe}$, of our Galaxy of (baryonic) mass $M_h$ can simply be related to the cumulative number, $N (X_{\rm Fe})$, of the stars born before the metallicity reaches $X_{\rm Fe}$ as; 
\begin{equation}
M_h X_{\rm Fe} = \langle Y_{\rm Fe} \rangle N (X_{\rm Fe}) f_{\rm SN}.  
\label{eq:yieldchem}
\end{equation}
   where $\langle Y_{\rm Fe} \rangle$ is the averaged iron yield per supernova and $f_{\rm SN}$ is the fraction of EMP stars that have exploded as supernovae, defined in eq.~(\ref{eq:frac-SN}). 
   By differentiating it with respect to $\feoh = \log (X_{\rm Fe}/ X_{\rm Fe, \odot})$, the number distribution of EMP survivors is written as a function of metallicity in the form
\begin{equation}
  n (\feoh) = \frac{ d N (X_{\rm Fe})}{ d\feoh} = \frac{M_h} {\langle Y_{\rm Fe} \rangle f_{\rm SN}} \ln(10) X_{{\rm Fe}\odot} 10^{\feoh} .
\end{equation}
   This shows that the number distribution of EMP survivors is simply proportional to the iron abundance apart from the variation of ${\langle Y_{\rm Fe} \rangle}$ through the IMF and the latter is small enough to be neglected for $\mmd \lesssim 20 \msun$ (see Fig.~\ref{fig:yield} in Appendix). 

\begin{figure}
\epsscale{1.0} 
\plotone{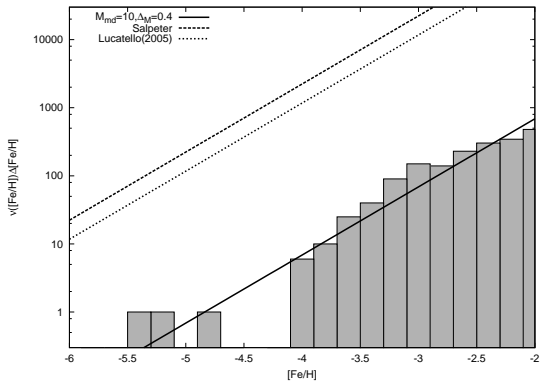}
\caption{Comparison of the theoretically predicted MDF from the IMF derived from the statistics of EMP stars (solid line) with the observed MDF obtained by the HES survey \citep[shaded columns;][]{Beers05b}. 
   The number distribution, $\nu_{\rm surv}$ is computed with the log-normal IMF of medium mass $\mmd=10 \msun$ and dispersion $\dm = 0.4$ with the 50\% binary fraction under the same flux-limited condition as the HES survey.  
   The examples of low-mass IMFs are shown for the Salpeter's power-law mass function (dashed line) and for the log-normal IMF ($\mmd = 0.79 \msun$ and $\dm = 1.18$, dotted line), derived by \citet{Lucatello05} from the \cemps\ star statistics alone, which bring about the overproduction of low-mass survivors.  
}
\label{MDF}
\end{figure} 

Figure~\ref{MDF} depicts the number distribution of EMP survivors and compares it with the observed MDF provided by the HES survey \citep{Beers05b}. 
	We assume stars of mass $m>M_{up} =8\msun$ become type II supernova and eject $\langle Y_{\rm Fe} \rangle =0.07\msun$ of iron. 
   In this figure, the theoretical MDF, $\nu_{\rm surv}$, is evaluated under the same flux-limited condition as the observed MDF is derived; 
   $\nu_{\rm surv} (\feoh) = n (\feoh) f_{\rm G} \times (40 \%) \times (8225 \hbox{ degree}^2 / 4 \pi \hbox{ sr}) \times 1.93$.  
   Here the fraction of follow-up observation and the sky coverage are taken into account: 
   as for the contribution of TO stars, we take the same ratio to the giants as in the observed sample under the assumption that the giant survivors are all reached in the survey area.  
   Solid line shows the MDF for the IMF with $ \mmd = 10 \msun$ and $\dm =0.4$ with the 50\% binary fraction, derived above for EMP population stars for the flat mass-ratio distribution, and it is similar to the other mass-ratio functions, as discussed in \S 2.3.  
   This reasonably reproduces the observed MDF between the metallicity $-4 \lesssim \feoh \lesssim -2.5$, as expected from the discussion in the previous section.  

In this figure, we also plot the MDF using the low-mass IMFs, the Salpeter's power-law mass-function as observed among the present-day stellar populations and that derived only from the statistics of \cemps\ stars by \citet[][$\mmd = 0.79 \msun$ and $\dm=1.18/\ln 10$]{Lucatello05}.  
   They bring about the overproduction of EMP survivors by a factor of more than a few hundreds not only from the low-mass members of binaries but also from the primary stars and the single stars; 
   both the IMFs give the similar MDF since our flux-limited samples are dominated by the giants and luminous dwarfs of mass $M \simeq 0.8 \msun$. 
   This means that the EMP survivors is by far a small population as compared with the stellar systems of Pop.~I and II, and it is only with the high-mass IMFs that can make the EMP population produce sufficient amount of metals to enrich the early Galactic halo without leaving too many low-mass survivors now observable in Galactic halo. 

In addition, we see in this figure that the slope of observed MDF is consistent with the prediction from the simple one-zone approximation at least for $\feoh > -4$.
   It implies that the IMFs have little changed while the Galactic halo has evolved through these metallicities.  
   Beyond $\feoh \simeq -2$, the observed MDF derived from the HK and HES surveys seems to be underestimated since those objects are out of the metallicity range sought after by the survey and subject to imperfect selection.  

\subsection{Effect of Hierarchical Galaxy Formation}

The observed MDF of Galactic halo stars has a sudden drop at $ \feoh \lesssim -4$, and only three stars are found below it\footnote{There are two more stars with the iron abundances reported below $\feoh<-4$; 
   CD $-38^{\circ}245$ with $\feoh = -4.19 \pm 0.10$ \citep{Cayrel04} and G77-61 with $\feoh = -4.03 \pm 0.1$ \citep{Plez05}. 
   We will omit these two stars in our discussion since larger abundances of $\feoh = -4.07 \pm 0.15$ \citep{Francois03} and $-3.98 \pm 0.15$ \citep{Norris01} have been reported for the former, and hence, their abundances are closer to the EMP stars of $\feoh \gtrsim -4$ than to the other three HMP/UMP stars. }. 
   We propose the mechanism responsible for this depression of low-metallicity stars from the consideration of the Galaxy formation process. 

In the current cold dark matter (CMD) model, galaxies were formed hierarchically.  
   They started from low mass structures and grew in mass through merging and accreting matter, finally to be large-scale structures like our Galaxy. 
   In the hierarchical structure formation scenario with $\Lambda$CDM cosmology, the typical mass of first star forming halos is $\sim 10^6 \msun$ for the dark matter and $\sim 2 \times 10^5 \msun$ for the gas \citep[e.g., see][]{Tegmark97,Spergel07}. 

In these first collapsed gas clouds, the first stars contain no pristine metals except for lithium. 
   When the first star explodes as supernova, it ejects $\sim 0.07 \msun$ of iron, which enriches the gas cloud of mass $\sim 2 \times 10^5 \msun$ where it was born up to the metallicity of $\feoh \sim -3.5$ if the ejecta is well mixed in the gas cloud.  
   We call this event the ``first pollution".  
   Consequently, the 2nd generation stars have the metallicity of $\feoh \sim -3.5$.

In the course of time, the mini-halos that host the gas clouds merge with each other and accrete the intergalactic gas to form early Galactic halo with the baryonic mass of $10^{11} \msun$. 
   We may take the metallicity of this early Galactic halo to be $\feoh \simeq -4$ because of the scarcity of stars of metallicity $\feoh<-4$.  
   The cumulative number of stars born before the early Galactic halo is enriched up to $\feoh=-4$ is estimated at $ N(10^{-4} X_{{\rm Fe} \odot}) = 7 \times 10^5 $ with taking into account the supernova fraction $f_{\rm SN}$ .  
   If the mini-halos of larger masses stand between the first collapsed halos and the Galactic halo, the dilution of iron with unpolluted primordial gas can give birth to the stars of smaller metallicity of $\feoh \simeq -4$, and then, the metallicity at the formation of Galactic halo can be larger to increase the cumulative number of stars in accordance (see below).  

We may estimate the fractions of both the first generation stars without metals and the 2nd generation stars of the metallicity $\feoh \sim -3.5$, respectively, assuming that stars are born with an equal probability whether in the gas clouds, polluted with metals, or in the primordial gas clouds.  
\red{   Here it is worth noting that some recent computations of star formation demonstrate that the low-mass stars can be formed as the binary members even out of metal-free gas \citep{Clark08,Machida08b}.  }
   Accumulated number, $N_{\rm Pop III}$, of Pop~III stars, born of gas unpolluted by SN ejecta, when the average metallicity reaches $X_{\rm Fe}$, is given by 
\begin{equation}
N_{\rm PopIII} = \frac{M_h}{M_c f_{\rm SN}} \left[1 - \exp \left(-\frac{M_c X_{\rm Fe}}{\langle Y_{\rm Fe} \rangle }  \right)\right] , 
\label{eq:num-popIII}
\end{equation}
   where $M_c (=2 \times 10^5 \msun)$ is the mass of gas in the first star forming clouds. 
   If we assume the same IMF and binary parameters as in the stars of EMP population, then, we expect that the number of Pop~III stars is 
\begin{equation}
N_{\rm PopIII} (10^{-4} X_{\rm Fe, \odot}) = 3.1 \times 10^5, 
\end{equation}
and the number of Pop~III survivors is 
\begin{eqnarray}
&&3.1 \times 10^5 \times \int_{0.08 \msun}^{0.8\msun} dm \big[\xi(m) + f_b \int_{0.8\msun} n (m/m_1) \xi(m_1) \frac{d m_1}{m_1} \big] \nonumber \\
&&= 1.3 \times 10^4,   
\end{eqnarray}
   and similarly we have $5.5 \times 10^4 $ and $2.3 \times 10^3 $ of the 2nd generation stars and their survivors, formed before the averaged metallicity of the Galaxy reaches $\feoh =-4$. 
   The IMF of Pop.~III stars may differ from EMP stars but the existence of the stars with $\feoh<-5$ suggests that the low-mass stars can be formed before the first pollution. 

Figure~\ref{MDFpop3} illustrates an expected MDF with the hierarchical structure formation.  
   After the formation of large Galactic halo, the metal enrichment process is thought to follow the argument of the previous subsection. 
   Thus, we can explain the cutoff around $\feoh \sim -4$ naturally. 
   Shaded columns indicate the initial distributions of Pop.~III stars and of the 2nd generation stars formed in the low-mass clouds. 
   The 2nd stars were mixed and observationally lost their identities among the stars formed in the merged halo. 
   On the other hand, Pop.~III stars should form a distinctive class.  
   From the above estimate, we expect $\sim 23$ Pop.~III survivors in the existing flux-limited samples of HES surveys. 
   \red{It is true, however, that there is no star with zero metallicity among the stars detected by the existent surveys. 
   We may propose one scenario to explain this absence that Pop.~III survivors are no longer remain metal-free at present since their surface are polluted by the accretion of interstellar matter, enriched with metals ejected by the supernovae of the first and subsequent generations. 
   With the surface pollution of $\feoh \sim -5$, they are observed as HMP stars. 
   We discuss about evolution of Pop.~III stars with pollution in \S \ref{HMPsec}. }
   Similarly, the number of second-generation of stars is estimated at $\sim 4$ in the same flux-limited HES sample, indicative that most of EMP stars are formed of mixture of the ejecta from plural supernovae.  
   This has direct relevance to the study of the nucleosynthetic signatures on the EMP survivors and the imprints of supernovae of the first and subsequent generations.   


\section{Conclusions and Discussion}

We have studied the initial mass function (IMF) and \red{the low-mass star formation with the} chemical evolution of the Galactic halo population on the basis of the characteristics of extremely metal-poor (EMP) stars, revealed by the recent large-scaled HK and HES surveys; 
   the observational facts that we make use of are; 
   (1) the overabundance of carbon-enhanced EMP (CEMP) stars, 
   (2) the relative frequencies of CEMP stars with and without the enrichment of s-process elements, 
   (3) the estimate of surface density or total number of EMP stars in the Galactic halo, and
   (4) the metallicity distribution function (MDF).  
   We take into account the contribution of binary stars properly, as expected from the younger populations.  
   In Paper~I, the high mass IMF peaking around $\sim 10 \msun$ is derived for the stars of EMP population and it is shown that the binary population plays a major role in producing the low-mass stars that survive to date, but by using the flat mass-ratio distribution between the component stars. 
   In this paper, we examine these properties of the stars of EMP population and EMP survivors for the different types of mass-ratio distributions and investigate the constraints on the IMFs of the stars of EMP population and discuss the observational tests of discriminating them.  
   The derived IMFs are applied to understand the characteristics of MDF and the nature of EMP stars including HMP/UMP stars, provided by the surveys.  

   Our main conclusions are summarized as follows; 

\noindent (1)  The statistics of CEMP stars are explained by the high-mass IMFs with the binaries of significant fraction.  
   Predicted typical mass is significantly larger than Population~I or II stars, \red{irrespective of the assumptions of the mass-ratio distribution. }  
\blue{   The mass-ratio distribution with a preference for nearby equal masses demands the IMF with higher typical mass $\mmd > 7\msun$ and smaller dispersion ($\dm<0.6$).  
   While the mass-ratio distribution in favor of smaller mass secondary or the independent combination of two stars with the same IMF demands smaller typical masses around $\mmd \sim 5 \msun$ irrespective of $\dm$. }

\noindent (2) \blue{High mass IMFs with $\mmd \sim 5-20 \msun$ derived from the statistics of CEMP stars agree with those derived from the low-mass star formation and the chemical evolution of Galactic halo based on the number of giant EMP sourvivors evaluated from the surveys. 
   IMFs with $\mmd \gg 20\msun$ are excluded by overproduction of iron or underproduction of EMP survivors.  
   IMFs with $\mmd \ll 5\msun$ need other iron source(s) without producing low-mass stars. }

\noindent (3) The mass-ratio distribution of binaries in the EMP population can be discriminated by the imprints left on the EMP survivors such as the mass function, the binary fraction, and the fraction of stars influenced by the supernova explosion of primary stars. 
\blue{   In particular, the flat mass-ratio distributions predict significant fraction ($\sim 7.7\% for \mmd=10$) of double-lined spectroscopic binaries while the mass-ratio distribution of ``independent'' coupling predict much lower fraction.  
   Among the 39 unevolved stars of $\feoh < -3$, studied spectroscopically to date, three double-lined binaries are found, but there may be significant uncertainties and biases for the existent surveys and future observations can discriminate the distributions. }

(4) The observed MDF of EMP survivors is consequent upon the derived IMF with the contribution of the binaries.  
   There is no indication of significant change in the IMFs between the metallicity of $-4 \lesssim \feoh \lesssim -2$.   
   The depression of stars below $\feoh < -4$ is naturally explicable within the current favored framework of the hierarchical structure formation model.  
   Then, the Pop.~III stars born of primordial gas, and also, the stars in the primordial clouds before they are contaminated by their own supernovae, should form the distinct class other than EMP stars, and may have the relevance to HMP and UMP stars observed at lower metallicity, as discussed below in this subsection.  

The feature of our approach is to take into account the stars born in binary systems properly in discussing the low-mass star formation in early Universe, based on the finding in Paper~I.    
   In addition, we make full use of available information from the existent large-scaled surveys and to draw the maximal constraint on the early evolution of our Galactic halo.  
   The known EMP stars ($\feoh \lesssim -2.5$) with the detailed stellar parameters amount to $\sim 400$ in number \citep[SAGA Database;][]{Suda08}, and allow us to discuss the averaged properties as studied in this paper.  
\green{   The existent surveys reach sufficiently deep and the nominal depth in the magnitude $12<B<17.5$ may corresponds to the heliocentric distance $d \simeq 10$ kpc or beyond for giants of $L \simeq 100 L_\odot$ while $d \simeq 0.3 -3$ kpc for dwarfs.   
   Incomplete coverage may affect the estimate of the total number of EMP survivors in the Galactic halo, and yet, the uncertainties will not be so large to exceed an order of magnitudes, judging from the density distribution of bright halo $\rho \sim r^{-3 \sim -3.5}$ \citep{Majewski93}.  
   Furthermore, because of strong dependence of the number ratio between the low-mass stars that survive to date to the massive stars that have exploded as supernovae on IMF, our} discussion in \S 2.3 through the iron production consistent with the number of EMP survivors will be left largely unaffected even quantitatively.  
   In order to improve and sharpen our conclusions, we have to wait for the future larger-scaled surveys such as SDSS/SEGUE \citep{Beers04} and LAMOST \citep{Zao06}.  
   Also the high dispersion spectroscopy is necessary to understand the characteristics of EMP stars.  

The constraints on the IMFs derived in this work may serve as the basis of understanding the formation and early evolution of the Galaxy. 
\green{   Although the HK and HES surveys may suffer from imperfect selections for $\feoh \gtrsim -2$, \citet{Ryan91} report that the metallicity distribution extends continuously from the peak at $ \feoh \simeq -1.6$ down to $\feoh \simeq -3$ without a break for the halo subdwarfs in the kinematically-selected samples. 
   This may be taken in turn as suggesting that the high-mass IMF derived for the EMP population persists without indication of significant changes up to such a large metallicity.  
   It forms a contrast with the prevalence of low-mass IMF with the characteristic mass $M \sim 0.3 \msun$ in the diverse conditions including the Galactic spheroid population of the metallicity $\feoh \simeq -1.7 - -1.4$ \citep{Chabrier03} and the globular cluster, while evidences, suggesting IMF biased toward the high masses, are reported from cosmological observations \citep[e.g., see][]{Elmegreen08}.  }
   Accordingly, there should be the transition from the high-mass IMF to the low-mass one.  
   Our result suggests that the transition is postponed until high metallicity even beyond $\feoh \simeq -2$ is reached.    
   It is likely that the transition may not be simply determined by the metallicity alone, 
\green{and may be related to the structure changes associated with the merging processes.  
   For the proper understandings of the transition, more theoretical works are necessary with the effects of hierarchical formation of Galactic structures properly taken into account. } 
   In discussing the primordial stars or HMP/UMP stars in the present work, we assume the metallicity at the formation of Galactic halo at $\feoh \simeq -4$.  
   The detailed chemical evolution with the merger history taken into account is discussed in a subsequent paper (Komiya et al.~ 2008, in preparation), in the similar ways as done by \citet{Tumlinson06} and by \citet{Salvadori07}, but taking into account the high-mass IMF, derived above, and the contribution of binaries.  
   
\subsection{Origin of HMP/UMP Stars}\label{HMPsec}

We end by discussing the consequences of the present study on the understanding of the origin of stars found below the cut-off of MDF.    

In our model, the stars made after the first pollution have the metallicity $\feoh \simeq -3.5$ and the stars with slightly lower metallicity of $\feoh \simeq -3.5 - -4$ are made in the merged clouds where metals are diluted with the primordial gas unpolluted by supernova ejecta. 
   After the halos merge, the 2nd generation stars mingle and observationally lose their identities among the stars formed in the merged halo. 
\blue{   This means that the HMP and UMP stars are the stars formed before the first metal pollution by the type II supernova in their host mini-halos. 
   One possible scenario for these stars is that they are the survivors of Pop.~III stars and the metal abundances at their surface are influenced not only by the matter from their binary companions but also by the interstellar matter, enriched with metals ejected by the supernovae after their birth.  }
\red{   In fact, it is shown that their peculiar abundance patterns of light elements from Li, carbon through aluminum, including s-process elements such as Sc and Sr, observed for three known HMP/UMP stars can be reproduced by neutron-capture nucleosynthesis during the AGB phase of their primary stars even under the pristine metal-free condition \citep{Nishimura08}.  
   As for iron group elements (Ca - Zn),} \citet{Suda04} argue the effects of surface pollution of Pop.~III stars through the accretion of interstellar gas to show that the main-sequence Pop.~III stars can be polluted to be $\feoh \simeq -3$ while the giants to be $\feoh \simeq -5$ since the pollutant is diluted by the surface convection deepening $\sim 100 $ times in mass on the giant branch. 
   Thus, the Pop.~III survivors have evolved to giants to be observed as HMP/UMP stars.  
\red{   As for a sub-dwarf HMP star HE1327-2623, the dilution of the accreted iron group elements has occurred in the envelope of primary star on the AGB because of the low-mass nature of its primary star \citep[$M \lesssim 1.5 \msun$,][]{Nishimura08}.  }

\blue{Majority of the Pop.~III survivors have also to be the secondary members of binary systems similar to the EMP survivors if their IMF and binary parameters are similar to EMP stars. }
   Then some of Pop.~III stars become carbon-enriched HMP/UMP stars with $\feoh \sim -5$ through binary mass transfer.  
   If the mass of primary star is $0.8 \msun < m_1 < 3.5 \msun$ and $3.5 \msun \lesssim m_1 < M_{up}$, the primary star enhances the surface abundances of carbon and nitrogen though the He-FDDM and of carbon and/or nitrogen through TDU and hot bottom burning in the envelope, respectively, which are transferred onto the secondary stars through the wind accretion.  
   It is to be noted that the primary stars of $m_1 >2 \msun$ have the accreted pollutants mixed inward into the whole hydrogen-rich envelope at the second dredge-up, and thereafter, evolve like the stars with the pristine metals. 
   At the same time, the accreted matter is diluted in the envelope and the iron abundance is reduced to $\feoh \sim -5$ in the primary stars. 
   We estimate that $\sim 35 \%$ of Pop.~III stars become carbon-rich HMP/UMP stars under the same assumptions on the binary parameters as in Paper~I.  
\blue{   The surface abundances of main sequence stars can be smaller than stated in Paper~I, however, since the accreted matter mixes down and are reduced by an order of magnitude if the diffusion and thermohaline mixing works \citep{Weiss00,Stancliffe07}. }

\begin{figure}
\epsscale{1.0} \plotone{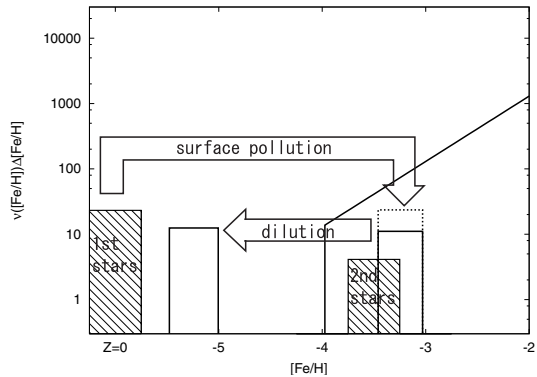}
\caption{Schematic drawing of MDF of EMP and Pop.~III survivors, constructed based on the hierarchical scenario for structure formation.
 Primordial main sequence stars with $Z=0$ polluted through the after-birth accretion of the interstellar gas, enriched with iron ejected by supernovae of the first and subsequent generations, upto $\feoh\sim -3$, and pollutants is diluted to $\feoh\sim -5$ in the surface convection of EMP survivors as it develops during ascent of red giant branch. 
}
\label{MDFpop3}
\end{figure} 

In Fig.~\ref{MDFpop3}, solid lines denote the expected MDF at the present days with the surface pollution taken into account.  
   The basic form of observed MDF is reproduced, i.e., the cutoff around $\feoh \sim -4$, the scarcity of stars for the metallicity below it and the existence of a few HMP/UMP stars.  
   From the above estimates, there should be $\sim 23$ Pop.~III stars in the existent flux-limited samples of HES surveys; 
   about a half of them may be discovered as giants with the surface metal pollution and one third as carbon stars. 
   \red{   In actuality, only three HMP/UMP stars (two giants and one sub-dwarf) are found to date, all enriched with carbon, among 153 stars of $\feoh <-3$, registered in SAGA database \citep{Suda08}. 
   Since such low metal abundances can be discriminated only with high dispersion spectroscopy, $\sim 4.6$ HMP/UMP stars are expected among the whole HES samples of 234 stars of $\feoh <-3$.  
   The observed numbers are significantly smaller than prediction from our model.  
　　} 
   The above estimates are made, however, under the assumption that the Pop.~III stars are formed in the same IMF as EMP stars and with the same binary parameters. 
   This may not be warranted and rather we may take that this deficiency may suggest a still higher-mass IMF and/or less efficiency of binary formation for Pop.~III stars than the EMP stars.  

In the above discussion, we assume the closed box chemistry in the collapsed object before merging. 
   It is shown that the hypernovae, exploded with a large energy of $10^{52}$ erg, blow off the first collapsed objects of mass $M \simeq 10^6 \msun$ \citep{Machida05}; 
   if the first stars are sufficiently massive, the metal yields are spread into larger masses, and pollute the ambient gas before they collapse to form mini-haloes, as discussed by \citet{Salvadori07}.  
   After that, the first stars in the collapsed clouds are no longer metal-free. 
   Nevertheless, those stars which are formed before each collapsed clouds are polluted by their own supernova form a distinct class from those which suffer from the first pollution.  
   Further study is necessary to make clear the present appearance of the possible Pop~III survivors and to settle the origin of HMP/UMP stars, in particular, for tiny amounts of iron-group metals and the overwhelming carbon-enhancement, shared by all these stars known to date.  

We benefit greatly from discussion with Dr.\ W.\ Aoki. 
This paper is supported in part by Grant-in-Aid for Scientific Research from Japan Society for the Promotion of Science (grant 18104003 and 18072001).
 
\appendix

\section{Low-mass star formation and iron production by EMP population}

One of the important findings of the recent large-scaled surveys is the scarcity of EMP stars in the Galactic Halo. 
   The HES survey gives the total number of EMP stars in our Galactic Halo at $\sigma _{\rm EMP} \simeq 796 \hbox{ sr}^{-1}$ (giants of $\sigma _{\rm EMP, G} \simeq 412 \hbox{ sr}^{-1}$ in eq.~(\ref{eq:HES-G}) plus turn-off stars $\sigma _{\rm EMP, TO} \simeq 384 \hbox{ sr}^{-1})$ within the limiting magnitude $B \simeq 17.5$. 
  Similarly, the HK survey gives $\sigma _{\rm EMP} \simeq 528 \hbox{ sr}^{-1}$ within the limiting magnitude of $B \simeq 15.5$; 
   114 stars of $\feoh < -3$ are found by the medium-resolution, follow-up spectroscopy of 50\% of the candidates, selected from the objective prism survey covering the $2800 \hbox{ deg}^2$ and $4100 \hbox{ deg}^2$ areas in the Northern and Southern Hemisphere \citep{Beers05a}. 
   Because of the significantly large areas covered by these surveys ($\sim 20\%$ of all sky with the follow-up observations), we may place reliance on these results, granted that they may not be complete. 
   This also constrains on the IMF of stellar population that promoted the chemical evolution, or more specifically, the formation of metals and the low-mass survivors.   
   In the paper, we have discussed the chemical evolution starting with the statistics of CEMP stars.   
   In this Appendix, we show that the chemical evolution with the total number of EMP survivors provides more stringent constraints on the IMF of EMP population with the aid of the amount of ejecta from supernova models, independently of the statistics of CEMP stars.  

Our basic premise is that the same stellar population is responsible both for the production of metals and of low-mass survivors. 
   In discussing the low-mass survivors, it is indispensable to take into account the contribution from the binaries. This is one of the major conclusions in Paper~I.  
   We assume that the stars are born not only as single stars but also as the members of binaries in an equal number and with the primary stars in the same IMF as the single stars.  
   For a given IMF, then, the total number, $N_{\rm EMP, surv}$ of EMP survivors, currently observed in the Galactic halo, is related to the cumulative number, $N_{\rm EMP}$, of stars of EMP population as;
\begin{eqnarray}
N_{\rm EMP, surv} & = & N_{\rm EMP } f_{\rm surv} \cr
& = & N_{\rm EMP } \int^{0.8 \msun} dm [ \xi(m) + f_b \int _{0.8 \msun}^{m_1} n(m /m_1) {d m_1 \over m_1}], 
\label{eq:frac-surv}
\end{eqnarray}  
   and hence, to the cumulative number of EMP supernovae as $N_{\rm EMP, SN}=N_{\rm EMP} f_{\rm SN} = N_{\rm EMP, surv} (f_{\rm SN}/ f_{\rm surv})$. 
   These supernovae have to supply the amount of iron, $M_{\rm Fe, EMP}$ in eq.~(\ref{eq:EMP-ironprod}), in order to enrich the gas in the Galaxy of mass $M_h$ with iron to promote the chemical evolution up to the metallicity $\feoh = -2.5$.  
   Then, we may derive the averaged iron yield, $\langle Y_{\rm Fe} \rangle_{\rm EMP}$, per supernova of EMP population, necessary to explain the chemical evolution of Galaxy, by the relation  $\langle Y_{\rm Fe} \rangle_{\rm EMP} = M_{\rm Fe, EMP} / N_{\rm EMP,SN}$ for an assumed IMF with the mass-ratio distribution function.  

\begin{figure}
\epsscale{0.5} 
\plotone{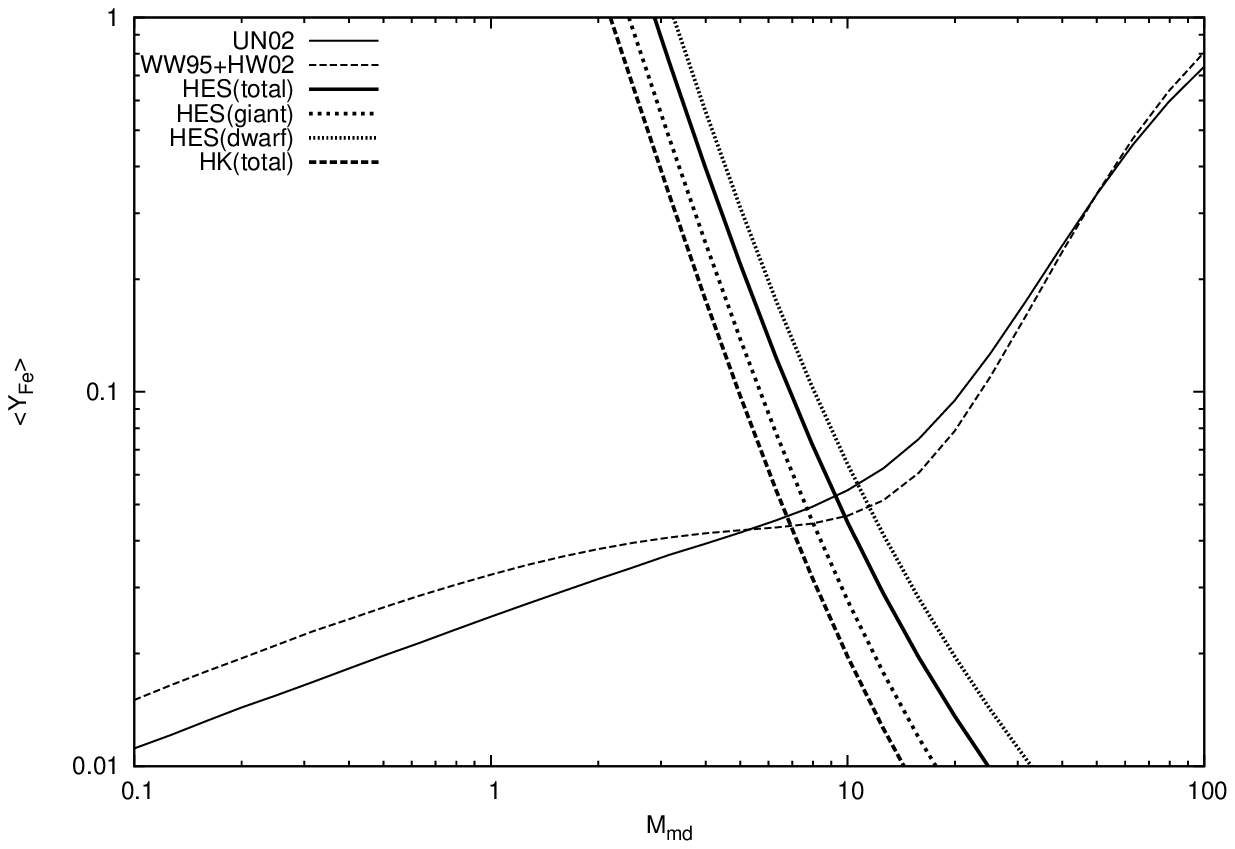}
\plotone{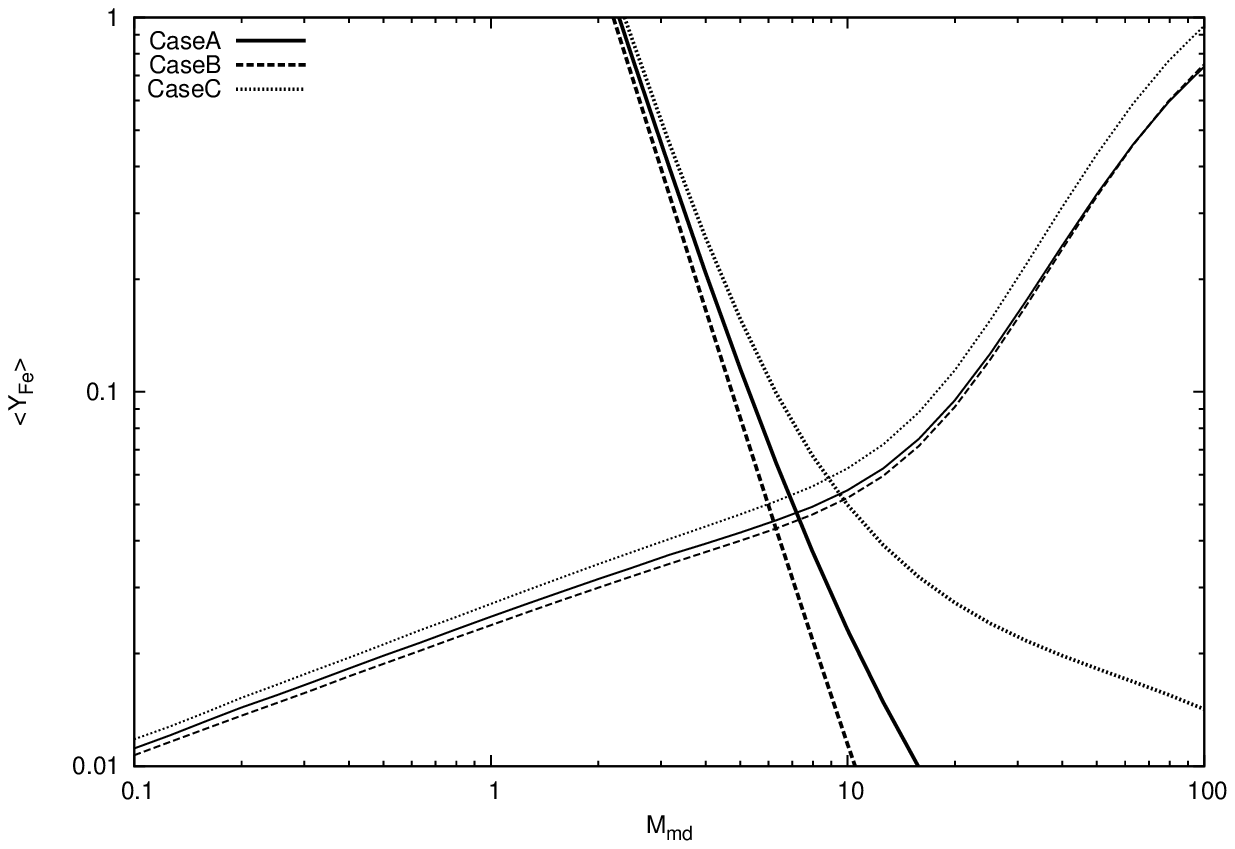}
%
\caption{The average iron yields per supernova, $\langle Y_{\rm Fe}\rangle_{\rm EMP}$, demanded from the chemical evolution of Galactic halo that leaves the EMP survivors consistent with the observed flux-limited samples, and the theoretical iron yields, $\langle Y_{\rm Fe}\rangle_{\rm SN}$, computed from the theoretical supernova models with use of ion yields by \citet{Umeda02}, and by \citet{Woosley95} and \citet{Heger02} as a function of $\mmd$ with $\dm = 0.4$.   
   Top panel compares $\langle Y_{\rm Fe}\rangle_{\rm EMP}$ computed for the different samples of EMP stars, i.e., giants, dwarfs and total stars from the HES survey and total stars from HK survey, with the flat mass ratio function, while the bottom panel compares the results with the different mass-ratio distributions for the giant samples of the HES survey.  
}
\label{fig:yield}
\end{figure}

We show in Figure~\ref{fig:yield} the averaged yield, $\langle Y_{\rm Fe} \rangle_{\rm EMP}$, as a function of $\mmd$ for $\dm = 0.4$:  
   upper panel for the observations of EMP stars of different evolutionary stages from the HES survey and of the total EMP stars from the HK survey with use of the IMFs with the flat mass-ratio function, and lower panel for the different mass-ratio functions with use of the observation of EMP giants from the HES survey. 
   In order to compare the stars of different evolutionary stages, we include the effects of the limiting magnitude of the surveys by assuming the de Vaucouleurs density distribution, $\rho\propto \exp(-r^{1/4})$, with the radial distance, $r$, from the Galactic center, the same as the stars in the Galactic halo and by assigning the luminosity of $L=L_\odot (M/\msun)^{3.5}$ and $100 L_\odot$ to dwarfs and giants, respectively.  

The amount of iron demanded by the chemical evolution turns out to be a steep decrease function of $\mmd$ since in order to leave a fixed number of low-mass survivors, the total number of stars of EMP populations, and hence, the supernova fraction increase rapidly with $\mmd$ in particular near $\mmd \simeq M_{up}$.  
   The necessary yields computed from the different samples in upper panel show a fairly good agreement with each other.  
   The difference between the giants and dwarfs for the HES samples is indicative of a relatively deficiency of dwarf stars compared with giants by a factor of $\sim 2.2$ in number, which may be attributed to rather crude assignment of averaged giant luminosity, and/or to the different efficiency of identifying giants and turn-off stars in the survey plates, and/or to the uncertainties in the spatial density distribution.  
   The results for the HK survey and the HES survey also agree within the difference by a factor of $\sim 2.3$ in number despite the difference in the limiting flux by 2 mag, and hence, to the difference in the searched volume by a factor of $\sim 20$.

The variations with the mass-ratio functions in the lower panel are caused by the difference in the number of supernovae per EMP survivor. 
   As compared to the flat mass-ratio function, the mass-ratio function increasing (decreasing) with $q$ give a larger (smaller) number of supernovae to produce one EMP survivors; 
   the difference of which increases for higher-mass IMFs. 

These iron yields necessary to promote the chemical evolution may be compared with the theoretical iron yields predicted from the supernova models.  
   The IMF-weighted iron yields, $\langle Y_{\rm Fe} \rangle_{\rm SN}$, per supernova is given by using the iron mass, $Y_{\rm Fe} (m)$, ejected from a massive star of initial mass $m$ as; 
\begin{equation}
\langle Y_{\rm Fe} \rangle_{\rm SN} = \frac{ \int_{M_{up} } dm_1 \xi (m_1) [ Y_{\rm Fe} (m_1)+ f_b \int_{M_{up}/m_1}^1 Y_{\rm Fe} (m_2) n(q) d q ]}{ \int_{M_{up} } dm_1 \xi (m_1) [ 1 + f_b \int_{M_{up}/m_1}^1 n(q) d q ]}.  
\label{eq:yieldsn}
\end{equation}
   The IMF averaged yield $\langle Y_{\rm Fe} \rangle_{\rm SN}$ is also shown in this figure, for which the theoretical yields are taken from the metal-deficient supernova models computed by \citet{Umeda02}, and by \citet{Woosley95} and \citet{Heger02}.  
   It is a slowly increase function of $\mmd$ for $\mmd \lesssim 20 \msun$ with the increase in the fraction of more massive stars that ended as supernovae, while beyond it, the gradient grows steeper owing to the contribution of the electron pair-instability supernovae of $M > 100 \msun$.  

The averaged yields, demanded by the chemical evolution, and the theoretical IMF-weighted iron yields both meet with each other near $\mmd \simeq 6-11 \msun$ and with the iron yield $\langle Y_{\rm Fe} \rangle \simeq 0.04-0.06 \msun$ per supernova.  
   As typically seen for the flat mass-ratio distribution, the parameter range coincides with that we have derived for the IMFs from the CEMP statistics in Figs.~\ref{fig:psi}-\ref{fig:nos}. 
  For higher-mass IMFs, the EMP stellar population cannot produce the sufficient number of low-mass survivors by themselves, while for lower-mass IMFs, it results short of iron production.  
   The differences arising from the mass-ratio distributions seem discernible but not large enough to differentiate these mass-ratio distributions in view of the uncertainties of current observations. 
   As compared with the flat distribution, the mass-ratio distribution increasing (decreasing) with $q$ demands smaller (larger) $\mmd$, the opposite tendency derived from the CEMP statistics.   
   These distributions prefer smaller (larger) number of EMP survivors, larger (smaller) fraction of \cemps\ stars and smaller (larger) ratio of \cempnos\ to \cemps\ stars.  
   In principle, however, we can discriminate the mass-ratio functions in the EMP binaries, including those destructed already by the evolution, with use of the survey and observations of EMP stars in sufficiently large number and with sufficient accuracy, which waits for future works.  
     
\green{In summary, the observed surface density of EMP stars indicates the high-mass IMF for the stars of early stage of Galactic evolution independently of the CEMP star statistics, and also indifferently of the assumed mass ratio distribution function.  
   It is true that the current estimate of $N_{\rm EMP /, surv}$ may be subject to significant errors, and yet, this result is robust because of a strong dependence of $\langle Y_{\rm Fe}$ on $\mmd$, as seen from the figure, which gives $ \delta  \log \mmd \simeq 0.4 \log N_{\rm EMP \, surv}$. }
   In the above discussion, we assume a single log-normal IMF with the binary fraction for the stellar population.  
   It is possible to assume the bi-modal IMF and to explain the production of iron and the formation of low-mass stars, separately,  in terms of the combination of two stellar populations, one with a higher-mass IMF responsible for the iron production and the other with a lower-mass IMF for the low-mass survivors, respectively.  
   In the case of bi-modal IMFs, the constraints, derived here, place an upper mass limit to the IMF of lower-mass population and an lower mass limit to the IMF of higher-mass population.  
   It is to be noted that the IMF with the binary mass function of Cases A-C is regarded as a sort of bi-modal IMF with the primary plus single stars as the higher-mass population and the secondary stars as the lower-mass population (see Fig.~12 in Paper~I); 
   the separation of two IMFs differs with the mass-ratio function and the relative contributions of two populations vary with the binary fraction.  
   In any case, as for the EMP stars in the Galactic halo, the statistics of CEMP stars, in particular, the ratio of \cempnos\ and \cemps\ stars, place the lower mass limit to the IMF of the lower-mass population, and hence, endorses the high-mass IMF, which narrows, if any, the contribution of the higher-mass population.

\end{document}